\documentclass[ 11pt]{article}
\pdfoutput=1
\usepackage{myoptions}
\usepackage{amsfonts}
\usepackage{amsmath}
\usepackage{url}
\usepackage{hyperref}

  \newlength{\abstractwidth}
\def\spc{\hspace{.5pt}}

\newcommand{\lrpar}[1]{\left( #1 \right)}
\newcommand{\corr}[1]{\left\langle #1 \right\rangle}

\def\hamgrav{H}

\def\GN{G}

\newcommand{\deltadelta}[2]{\frac{\delta #1}{\delta #2}}

\def\rrc{r_c} 
\def\rrct{r_c^2}
\begin{document}

\renewcommand{\footnotesize}{\small}
\numberwithin{equation}{section}

\begin{titlepage}
  \bigskip

  \bigskip\bigskip

  \bigskip

\begin{center}
{\Large \bf 
Moving the CFT into the bulk with  $T\bar{T}$}
 \bigskip
{\Large \bf { }} 
    \bigskip
\bigskip
\end{center}

  \begin{center}

 \bf {Lauren McGough,$^1$ M\'ark Mezei,$^2$ and Herman Verlinde$^{1,2}$ }
  \bigskip \rm
\bigskip

   $^1$Department of Physics, Princeton University, Princeton, NJ 08544\\
\rm
 \bigskip
 $^2$Princeton Center for Theoretical Science, Princeton University, Princeton, NJ 08544

  \bigskip \rm
\bigskip
 
\rm

\bigskip
\bigskip

  \end{center}

 \bigskip\bigskip
  \begin{abstract}
\noindent 
Recent work by Zamolodchikov and others has uncovered a solvable irrelevant deformation of general 2D CFTs, defined by turning on  the dimension 4 operator $T \bar T$,
the product of the left- and right-moving stress tensor. We propose that in the holographic dual, this deformation represents a geometric cutoff that removes the asymptotic region of  AdS and places the QFT on a Dirichlet wall at finite radial distance $r = \rrc$ in the bulk. As a quantitative check of the proposed duality, we compute the signal propagation speed, energy spectrum, and thermodynamic relations on both sides. In all cases, we obtain a precise match. We derive an exact RG flow equation for the metric dependence of the effective action of the $T \bar T$  deformed theory, and find that it coincides with the Hamilton-Jacobi equation that governs the radial evolution of the classical gravity action in AdS.

 \medskip
  \noindent
  \end{abstract}
\bigskip \bigskip \bigskip

  \end{titlepage}

\tableofcontents

\addtolength{\baselineskip}{0.4mm}

\addtolength{\parskip}{.75mm}

\addtolength{\abovedisplayskip}{1mm}
\addtolength{\belowdisplayskip}{1mm}

\def\fxi{f}

\pagebreak

\section{Introduction and Summary}

\vspace{-1mm}

AdS/CFT duality is a powerful statement, thanks to the fact that one partner in the duality is a manifestly well defined quantum  system with precise rules for computing correlation functions of local operators. Conformal field theory is by definition a UV complete framework, in which the rules of local quantum field theory apply at all energy scales. These statements remain true for relevant or marginal deformations of CFTs  that preserve the existence of a UV fixed point.

This virtue also has a flip side, as it makes AdS/CFT rather special.
CFTs, or more generally, quantum field theories that are connected via RG flow to a UV fixed point, form a set of measure zero within the space of all effective QFTs. It is then natural to ask: can  holography be extended to effective QFTs for which the UV behavior is {\it not} described by a CFT? In the context of AdS${}_3$/CFT${}_2$, this question has recently become more opportune, due to the discovery of Smirnov and Zamolodchikov \cite{smirzam} of a general class of exactly solvable irrelevant deformations of 2D CFT. Turning on an irrelevant coupling typically spoils the existence of a UV fixed point and destroys locality at some high cutoff scale. Properties of the deformed CFTs uncovered in \cite{smirzam}, however, are found to be robust and largely decoupled from the question of their UV completeness.

In this paper we consider the simplest example of a solvable irrelevant deformation a 2D CFT, obtained by turning on a $T \bar{T}$ coupling
\es{ttbardeform}{
S_\text{QFT} & \spc =\spc  S_\text{CFT}+\mu\int \! d^2x\ T\spc \bar T \,.
}Here $T \bar T$ denotes the composite irrelevant (dimension 4) operator given by the product of the left- and right-moving components $T\equiv T_{zz}$ and $\bar{T}\equiv T_{\bar{z}\bar{z}}$ of the stress tensor, where we defined $z= x+i\tau$. Note that because $T \bar T=\frac18\spc \le(T^{\al\beta}\spc T_{\al\beta}-(T^{\al}_{\al})^2\ri) $, the deformation preserves Lorentz invariance. By finite $\mu$ we mean that there is a one parameter family of theories defined by ${dS^{(\mu)}_\text{QFT}/ d\mu}=\int \! d^2x\ (T\spc \bar T)_\mu$, where the $\mu$ subscript of $T \bar T$ emphasizes  that in this equation we have to use the stress tensor of $S^{(\mu)}_\text{QFT}$. The deformation \eqref{ttbardeform} is exactly solvable, in the sense that, even if the original 2D CFT itself has no extra symmetries other than Virasoro symmetry, the deformed theory possesses an infinite set of conserved charges and allows for exact computation of interesting physical quantities such as scattering phases, energy levels, and the thermodynamic equation of state \cite{smirzam,italians,nyugroup}. 
Moreover, as we will see, there are several indications that the deformed CFT defined by \eqref{ttbardeform} represents a consistent unitary quantum theory, with many interesting properties that are worth exploring. 

We are interested in how the deformation \eqref{ttbardeform} affects the standard holographic dictionary \cite{Maldacena:1997re,Gubser:1998bc,Witten:1998qj} between CFT quantities and corresponding properties in AdS gravity . In the following we will argue that the coupling $\mu$ acts as a geometric cutoff that removes the asymptotic region of the AdS space-time, and thereby places the QFT at a finite radial distance $r\! =\! \rrc$ from the center of the bulk. We will test this proposal for the special subclass of quantities that can be created or measured by the stress  tensor, or equivalently,
by deformations of the metric.\footnote{The background metric in the presence of the $T \bar T$ deformation is defined via the relation $\< T_{\alpha\beta}\> = {2 \ov \sqrt{g}} {\delta  \ov {\delta g^{\alpha\beta}}}\log Z_\text{QFT}$, with $T_{\alpha\beta}$  the unique local conserved  current associated with translation symmetry. \label{foot1}}
 Examples of such quantities are signal propagation speeds, finite size effects, thermodynamic properties, and the Euclidean partition function $Z_\text{QFT}(g,  \mu)$ in a general background metric $ds^2 = {g}_{\alpha\beta} dx^\alpha dx^\beta$.

Our concrete proposal is that the {deformed} CFT \eqref{ttbardeform} is dual to the original gravitational theory (i.e.~the gravity dual of the original CFT) living on a compact sub-region of AdS space-time
\es{adsmet}{
\qquad\qquad ds^2_\text{AdS} &\spc  = \spc \spc {dr^2 \ov r^2}  \, + \, r^2 \, {g}_{\alpha\beta} dx^\alpha dx^\beta  \spc  ,
\qquad \ \ \ r < \rrc\,,
}
defined by restricting the radial coordinate to the finite interval $r<\rrc$,  with $\rrc$ related to   $\mu$ via
\es{rcutoff}{   \mu &\, =\,{16\pi \GN \ov \rrct}  \, = \, {24\pi \ov c } \spc {1\ov \rrct} \, .
}
Throughout the paper we set $\ell_\text{AdS}=1$, hence the Brown-Henneaux relation used in the above equation is $c={3\ov 2 G}$ \cite{Brown:1986nw}. At large central charge $c$, we can identify
\es{zaction}{ Z_\text{QFT}(g_{\alpha\beta},  \mu)& = \exp\Bigl({- \mbox{\Large $1\ov 16\pi G$} \, S_\text{cl}\bigl(\rrc^2\, g_{\alpha\beta}  \bigr)}\Bigr)\,,
}
where 
 $S_\text{cl}(\rrc^2 g_{\al\beta})$ is the classical action of the 3D gravity theory restricted to the region $r\!  <\! \rrc$, with Dirichlet boundary conditions $ds^2\vert_{r=r_c} = \rrct g_{\alpha\beta}dx^\alpha dx^\beta$ on the metric and $\phi_i\vert_{r=r_c} = 0$ on all bulk fields $\phi_i$. Here we assume that the classical matter fields do not contribute any stress-energy source.

The proposal has interesting implications  for the holographic renormalization group program. In the formulation of \cite{heemskerk} (see also \cite{boer-vv,Verlinde1999,liu,Harlow:2011ke}) the CFT partition sum $Z_\text{CFT}$ is identified with the gravity partition function in which the bulk path integral is cut into an IR and UV part via
\es{HoloRG}{
Z_\text{CFT}(\tilde{g}_{\al\beta}, \epsilon) &\spc  =\spc  \int {D g_{\al\beta}}\ 
\Psi_\text{IR}\bigl(r_c^2\spc  g_{\al\beta}\bigr)\, \spc \Psi_\text{UV}\le(r_c^2 \spc  g_{\al\beta},\, \epsilon^{-2}  \tilde{g}_{\al\beta}\ri)\,.
}
Here $\ep$ denotes the short distance cutoff of the CFT. Here we have suppressed the integral over all matter fields: we assume that their saddle point value can be consistently set to zero.
 $\Psi_\text{UV}$ is a path integral over metrics of the form \eqref{adsmet} over the region $\rrc< r < 1/\epsilon$ with prescribed boundary conditions, while $\Psi_\text{IR}$ is an integral over all metrics in the region $r<\rrc$ with boundary conditions that match those of $\Psi_\text{UV}$.  The IR wave-function $\Psi_\text{IR}$ satisfies the Wheeler-DeWitt constraints, and via the holographic dictionary,  is to be identified with a QFT path integral with a UV cutoff of size $1/r_c$.  The UV wave-function $\Psi_\text{UV}$ is related to the  Wilsonian action by an functional Legendre transform, and is local on distance scales larger than $1/r_c$. 
 
In this language our proposal states that 
\es{HoloRG2}{
 Z_\text{QFT}\le(g_{\al\beta}, \,  \mu  \ri)&=\Psi_\text{IR}\le(r_c^2\,  g_{\al\beta}\ri)\,
}
with $\mu$ and $r_c$ related via \eqref{rcutoff}. The full CFT partition function is insensitive to how we choose our renormalization scale, hence \eqref{HoloRG} is independent of $r_c$. Then the role of $\Psi_\text{UV}$ is to undo the $T\bar{T}$ deformation of the CFT to get back the CFT result for the full partition function. It is also important to note that if the CFT has a large $N$ counting, where $c=O(N^2)$, the $T\bar{T}$ deformation is an irrelevant double trace deformation. There has been earlier speculation that the sharp radial cutoff in the bulk could be related to this kind of deformations \cite{heemskerk}.

The proposed  dictionary is supported by several quantitative agreements between the two sides. We list three of them below.

 \smallskip

1. {\it Deformation of the light cone.} The physical consequences of the $T \bar T$ deformation become most apparent by considering the system at finite temperature or in some eigenstate with finite energy density. In both cases, the stress-energy tensor has a non-zero expectation value. 
As pointed out by Cardy \cite{cardy}, this leads to a renormalization of the propagation speed $v_\pm$ of left- and right-moving massless excitations. In Minkowski space \eqref{ttbardeform} takes the form
\es{ttbardeformMink}{
S_\text{QFT} & \spc =\spc S_\text{CFT}-\mu\int \! d^2x\ T_{++} \spc T_{--}\,, 
}
where we defined $x^{\pm}=t\pm x$.
Splitting off the expectation value  from $T_{++} $ and $T_{--}$, the deformed action \eqref{ttbardeformMink} acquires a linear term $-\mu \int d^2 x \ \le[\< T_{++} \>\spc\spc T_{--} \spc +\spc \< T_{--}\>\spc\spc T_{++} \ri]$, which has the same physical effect as a perturbation of the 2D metric of the form\footnote{To derive this equation, we used the definition $\de S_\text{QFT}=-\frac12\int d^2 x \ \de g^{\al\beta} \, T_{\al\beta}\,,$
and that $\de g^{\pm\pm}=-4\de g_{\mp\mp}$.
} 
$ds^2_\text{CFT}  \simeq - dx^+ dx^- -\spc {\mu\ov 2}\spc \<\spc T_{++}\spc \>\spc (dx^+)^2- {\mu \ov 2}\spc \<\spc  T_{--}\spc \>\spc (dx^-)^2\,.$
We see that the deformed CFT behaves like a gravitational theory in which stress-energy back reacts on the space-time geometry. The null directions of the effective metric are 
$dx^+=- {\mu\ov 2}\,  \<\spc  T_{--}\spc \>\spc dx^-$ or $dx^-=-{\mu \ov 2}\<\spc T_{++}\spc \>\spc dx^+\,,$ and the propagation speed for left- and right-movers thus gets renormalized~to
\es{newLC}{
v_\mp & \simeq \spc 1\spc +\spc {\mu}\, \< T_{\pm\pm} \>\,.
}
Note that for $\mu>0$, the deformation gives rise to superluminal propagation speeds, as the null energy $\< T_{\pm\pm} \>$ is non-negative in states to which the above hydrodynamic argument applies.

This effect has a natural interpretation in the gravity dual.  A high energy CFT state is dual to a BTZ black hole geometry \cite{Banados:1992wn}.   The propagation speed of metric perturbations of a BTZ black hole placed  with Dirichlet walls at $r\! =\! \rrc$  was analyzed by Marolf and Rangamani in \cite{marolf}. Somewhat surprisingly, they found that these perturbations propagate at superluminal speed relative to the metric at the cutoff surface.
Generalizing their derivation to the rotating case, one finds that the left- and right-moving propagation speeds are given by $v_\pm \simeq\spc 1+{(r_+ \mp r_-)^2\ov 2 r_c^2}$, with $r_+$ and $r_-$ the radius of the outer and inner horizon \cite{BTZ}. 
Equating the  renormalized velocities on both sides of the duality reproduces the standard result for the holographic stress-energy tensor \cite{holostress} in the BTZ background, provided that $\mu$ and $\rrc$ are related via \eqref{rcutoff}.

\smallskip

{\it 2. Deformed energy spectrum.} Another interesting physical quantity is the $\mu$ dependence of a given energy level $E_n(\mu, L)$ on a cylinder with circumference $L$.\footnote{In studies of 2D CFT on a cylinder, it is customary to set $L=2\pi$.\label{foot2}}  Remarkably,
this quantity can be computed exactly for any energy eigenstate of the perturbed CFT  \cite{smirzam,italians,nyugroup}. For a given CFT state with conformal dimension $(\Delta_n, \bar{\Delta}_n)$ one finds 
\es{energylevel}{
\qquad E_n(\mu, L) L& = \,  {2\pi\ov \tilde\mu}\, \le(\spc 1-\sqrt{1-2{\tilde\mu}\spc M_n +  \spc \tilde\mu^2\, J_n^2}\, \ri)\, ,  \qquad \ \tilde{\mu}\spc \equiv \spc {\pi \mu \ov L^2}\,,
}
with $M_n =  \Delta_n + \bar{\Delta}_n - {c \ov 12}$, and  $J_n =\Delta_n-\bar{\Delta}_n$. Note the right-hand side becomes imaginary above some critical conformal dimension (for fixed $\tilde{\mu} > 0$) or above some critical value of $\tilde{\mu}$ (for fixed  $\Delta_n + \bar{\Delta}_n> {c \ov 12}$). This behavior is called the `shock singularity' in \cite{smirzam} and indicates the presence of a UV cutoff.
We will summarize the derivation of the result \eqref{energylevel}  in section \ref{sec:spectrum}. 

The analogous quantity to $E_n(\mu,L)$ on the gravity side is the quasi-local energy of a BTZ black hole of mass $M$ and angular momentum $J$ placed in a spatial region $r<\rrc$,
with Dirichlet boundary conditions $ds^2\vert_{r=r_c} = r_c^2 \, dx^+dx^-$. This quantity was computed in \cite{brownmann} (shortly before the discovery of AdS/CFT, so without any reliance on or reference to the holographic dictionary) by integrating the Brown-York stress-energy tensor over the boundary surface. This gravity result, given in equation \eqref{BTZcutoff},
and the QFT result \eqref{energylevel} precisely match, again provided we identify $\mu = {24\pi \ov  c}{1\ov r_c^2}$.

The agreement between the energy spectra extends to a precise correspondence between all thermodynamic quantities, such as the equation of state, pressure, temperature, heat capacity, etc.  Since the equations remain valid for finite values of $\tilde{\mu}$, this provides a new tool for studying bulk physics deep inside AdS. In particular,  the `shock singularity' of the deformed CFT (above which $E_n(\mu, L)$ becomes imaginary) is mapped to the singular properties (such as a diverging temperature and pressure) of the BTZ black hole inside a box $r\! < \! \rrc$ as $\rrc$ approaches the horizon. Studying the nature of this transition may give new insight into the physics of black hole horizons.
 
\smallskip

{\it 3. Exact RG equation.}  A key property, on which many of the exact results about the deformed theory \eqref{ttbardeform} are based, is the following  relation 
 for the expectation value of the composite operator $T \bar T$
\es{zamttb}{
\< \spc T \spc \bar T \spc \> & = \< \spc T \spc \> \< \spc  \bar T \spc \> - \< \spc \Theta \spc \>^2\,.
} 
Here $\Theta = T_{z\bar{z}}=\frac14 T^\alpha_\alpha$ denotes the trace of the stress tensor.  This remarkable factorization property was first derived by Zamolodchikov in \cite{zamttb} and holds for any translation invariant, stationary state in any relativistic 2D QFT. Equation \eqref{zamttb}  in particular implies that the composite operator $T\bar T$ has exact scaling dimension 4, up to possible total derivative terms.\footnote{Typically, a factorization property of this type  is  only exact  in a strict large $N$ limit or for suitable protected operators in supersymmetric QFTs.} The absence of anomalous dimensions makes it possible that energy spectrum \eqref{energylevel} is independent of the UV cutoff.  \eqref{zamttb} can be used to derive an RG equation for the partition function of the deformed CFT as follows.

The partition function $Z_\text{QFT}(g, \mu)$ of the deformed CFT has a prescribed dependence on the 2D metric. Using that $\mu$ is the only scale in the problem,  by taking the functional derivative of the partition function with respect to the scale factor of the metric to first order in $\mu$ we get
\es{scaledeform}{
\< \spc \Theta \spc \>  & = - {c\ov 96\pi} R(g)- \spc  {\mu\ov 2} \spc \< \spc T \spc \bar T \spc \>\,.
}
The first term on the right-hand side is the trace anomaly of the CFT, the second term is a correction due to the $T\bar T$ deformation. Assuming that the metric is slowly varying, we combine \eqref{scaledeform} with the Zamolodchikov relation (\ref{zamttb}) to get
\es{exactrg}{ 
\< \spc \Theta \spc \>  & =  - {c\ov 96\pi} R(g) - \spc  {\mu\ov 2} \spc  \le( \< \spc T \spc \> \< \spc  \bar T \spc \> - \< \spc \Theta \spc \>^2\ri) \,.
}
The above equations \eqref{zamttb}, \eqref{scaledeform} and \eqref{exactrg} all hold to leading order in a derivative expansion. 

The result \eqref{exactrg} can be viewed as an exact RG equation of the $T \bar T$ deformed CFT. 
It holds for any 2D CFT, but acquires a special meaning for CFTs with gravity duals.
To make its interpretation more evident, let us insert the holographic dictionary \eqref{rcutoff} and \eqref{zaction} into \eqref{exactrg}. This leads to a non-linear first order differential equation for the classical gravity action, given in equation \eqref{HJeqn}, which
coincides with the Hamilton-Jacobi (HJ) form of the holographic RG equation \cite{heemskerk, 
boer-vv, liu} that governs the radial evolution of the classical gravity action in AdS  as a function of the cutoff $\rrc$. 

The results summarized above all have a common geometric origin.
The HJ equation  \eqref{HJeqn} is the classical limit of the Wheeler-DeWitt constraint that describes the radial evolution of a wave-function in 3D gravity.
It has been known for some time that the partition function of a 2D CFT can be mapped, via a simple integral transform  \cite{HV,freidel},  to a wave-function that solves the WDW constraint of 3D gravity. From the CFT perspective, this integral transform looks like the  $T \bar T$ deformation \eqref{ttbardeform}, rewritten  in terms of a Gaussian integral over metric fluctuations. This exact result,  stated in equations \eqref{legendre} and \eqref{freidel} in section \ref{conclusions},  provided the initial inspiration for our conjectured interpretation of the $T \bar T$ deformation as moving the CFT into the bulk.

\smallskip
\medskip

In the following sections we give some more detailed derivations of the above results. In section \ref{sec:TTbar}, we review the known exact results about the integrable $T \bar T$ deformation, including the presence of an infinite set of conserved charges, the energy spectrum \eqref{energylevel} and the Zamolodchikov equation \eqref{zamttb}. We also highlight a relationship between the $T \bar T$ deformation and the Nambu-Goto action. In section \ref{thermodynamics}, we review the computation of the quasi-local energy and thermodynamical properties of the rotating BTZ black holes with finite radial cutoff. 
 In section \ref{sec:speed} and the Appendix, we derive the renormalization of the propagation speed in CFT states dual to rotating BTZ black holes and a general class of bulk space-times giving space dependent stress tensor expectation values. In section \ref{sec:RG}, we look in more detail at the derivation of the exact RG equation \eqref{exactrg} and its relation with the WDW equations of the 3D gravity theory. We end in section \ref{conclusions} with a discussion of various open questions.

\def\LL{\mbox{\small $\Omega$\spc}}

\section{$T \bar T$ Deformed CFT} \label{sec:TTbar}

In this section we will give an overview of some exact properties of the $T \bar T$ deformed CFT. More details can be found in the original papers \cite{smirzam, italians, nyugroup}.
 These exact results are an important cornerstone of our general proposal. We will also address the question of UV completeness.
  As a concrete piece of evidence in favor, we point out that for the special case that the  CFT has central charge $c=24$, the $T \bar{T}$ deformation is exactly soluble and manifestly consistent --  and in fact equivalent to the worldsheet theory of critical string theory \cite{italians, nyugroup}.

\subsection{Integrability}

\vspace{-1mm}

The $T \bar{T}$ deformation is a special case of a more general class of irrelevant integrable deformations of CFTs introduced by Smirnov and Zamolodchikov (henceforth SZ) in \cite{smirzam}.
For completeness, we briefly state their main result. More details and derivations can be found in \cite{smirzam} (see also \cite{italians}).  Integrable deformations of 2D CFTs are characterized by the existence of an infinite set of conserved higher spin charges $P_s$ and $\bar{P}_s$ of the form 
\es{imsp}{
P_s & = \oint_C \! \bigl(\spc T_s\spc dz \spc+ \Theta_s \spc d\bar z\bigr) \qquad \ \ \bar{P}_s  = \oint_C \! \bigl(\spc \bar{T}_s\spc d\bar z \spc+ \bar{\Theta}_s \spc d z\bigr)\,
}
where the current components $T_s$ and $\Theta_s$ are local operators of spin $s+1$ and $s-1$, respectively, and satisfy the current conservation identity
\es{imcons}{
\p_{\bar z} T_s & = \p_z \Theta_s\,, \qquad \quad \p_z \bar{T}_s = \p_{\bar z} \bar\Theta_s\,.
}
The simplest integrals of motion with $s=1$ are the total left- and right-moving energy-momentum $P_+ = \oint_C (\spc T \spc dz \spc+ \Theta \spc d\bar z) $
and $P_- = \oint_C (\spc \bar{T} \spc d\bar z \spc+ \Theta \spc d z)$. 
The conserved charges $P_s$ all commute with each other by virtue of the fact that their commutator with the currents yields a total derivative.

In the undeformed CFT, all $\Theta_s = \bar \Theta_s =0$ and the currents $T_s$ and $\bar{T}_s$  are all chirally conserved. They are given by special  composite operators, generically made up from the left- or right-moving stress tensor, respectively.  For irrational CFTs, all currents $T_s$ and $\bar{T}_s$  are of this type. For CFTs with Kac-Moody or W-symmetries, there may be additional currents. For clarity, we emphasize that our notion of integrability does not automatically imply exact solvability: irrational  CFTs with holographic duals are typically not exactly soluble. However, thanks to the infinite Virasoro symmetry, they posses and infinite set of conserved charges, and allow for integrable deformations that preserve an infinite subset of them.

The main results of SZ is that, in the neighborhood of any 2D CFT within the space of all 2D QFTs, there exists an infinite parameter family of integrable QFTs obtained by turning on an infinite set of irrelevant deformations of the form  \es{imscft}{
S_\text{QFT} & = \spc S_\text{CFT} \spc +\spc  \sum_s \spc \mu_s\! \spc \int\! d^2x\spc X_s \,,
\qquad \qquad X_s  \,   \equiv \, \spc T_{s}\bar{T}_{s}- \Theta_{s} \bar\Theta_{s} \,. 
}
Since $X_s$ has scaling dimension $2s+2$, these theories all become strongly coupled in the UV.
Nonetheless, one can  derive exact results about their symmetries, integrability, scattering phases and energy spectrum. 
In particular, SZ show that the conserved charges $P_s$ can be defined such that 
${\p P_s  /\p \mu_{s'}} + \bigl[ P_s, \int \! d^2 x \, X_{s'}\bigr]  = 0$ for all $s$ and $s'$. This identity implies that all charges are preserved by the infinite set of deformations \eqref{imscft}.

Our interest  is in the special case that only the least irrelevant coupling $\mu=\mu_1$ of the lowest operator  $X_1 = T \bar T - \Theta^2$ is non-zero. The integrability of this deformation will not be central to our story, except that it helps with some exact computations and gives some confidence that the deformed theory is well defined.
Following SZ we often refer to the operator $X_1$ simply as $T \bar T$. 

\vspace{1mm}

\subsection{Zamolodchikov equation}
\vspace{-1mm}

The key result, from which many of the exact properties of the deformed CFT can be derived, is the Zamolodchikov equation \eqref{zamttb} for the expectation value of the composite operator $T \bar T$, which holds for any translation invariant state in any 2D QFT with a local stress tensor.  Here we briefly summarize its derivation \cite{zamttb}. Consider the difference of two point functions 
\es{genx}
{
\Xi(z,w) & \, \equiv \; \bigl\< T(z) \bar T(w) \bigr\> - \bigl\< \Theta(z) \Theta(w) \bigr\>\,.
}
By taking two opposite limits, this function $\Xi(z,w)$ formally reduces to the expectation value of the composite operator  $X_1 = T \bar T - \Theta^2$, or factorizes into the product of expectation values of individual stress tensor components
\es{limx}{
\lim_{w\to z} \Xi(z,w) & = \bigl\< T \bar T \bigr\> - \< \Theta^2 \bigr\> ,\\[2mm]
\lim_{w\to \infty} \Xi(z,w) & = \< T \> \<\bar T \> - \< \Theta \>^2 \,.
}
The second equality follows from the cluster property of local QFT. 
The first  limit a priori needs to be taken with care, since the OPE between two operators generally becomes singular at short distance.
The key insight, that relates the two limits and makes the first limit well behaved, is that the gradient of $\Xi(z,w)$ with respect to $z$ and $w$ identically vanishes. Using the conservation laws \eqref{imcons} and the fact that in a translation invariant state, the two point functions in  \eqref{genx}  depend only on the coordinate difference $z-w$, one readily derives that
\es{zamarg}{
 \bigl \< \p_{\bar z}T(z) \bar T(w) \bigr\> - \bigl\<\p_{\bar z}\Theta(z) \Theta(w)\bigr\> & =  - \bigl\<\Theta(z) \p_{w}\bar T(w) \bigr\> +\bigl \< \bar {T}(z)\p_{w} \Theta(w)\bigr\>= 0\,,
}
which shows that $\p_{\bar z} \spc\Xi(z,w) =0$. In a similar way, one derives that $\p_z \spc\Xi(z,w) =0$. Hence the function $\Xi(z,w)$ is a constant. This proves that the two right-hand sides in  \eqref{limx} are equal, leading to the relation 
\eqref{zamttb}. 

Let us make two cautionary comments. First, as mentioned in the introduction, the relation \eqref{zamttb}  suggests that the composite operator $T \bar T$ has exact scale dimension 4.  This seems a surprisingly strong statement. However, since the  derivation of \eqref{zamttb} makes essential use of translation invariance, this property only holds at zero momentum. A more cautious and correct statement is that $T \bar T$ behaves as a local scaling operator with scale dimension 4 up to total derivative terms \cite{zamttb}. In later sections, we will also make use of \eqref{zamttb} as a property that describes the behavior of more general states to leading order in a derivative expansion.
A second related point is that in what follows, we will assume the result \eqref{zamttb} remains valid for the $T \bar T$ deformed CFT at finite coupling $\mu$. Since the derivation outlined above applies to any QFT, this seems reasonable. However, as explained in \cite{zamttb}, to avoid possible ambiguities related to the total derivative terms may require extra assumptions about the UV behavior of the QFT, which may not  obviously hold for the $T \bar T$ deformed theory. We will ignore this subtlety in what follows.

\def\TT{ \mbox{\small $T$}}

\def\stars{{\mbox{\tiny $\text{H}$}}}
\subsection{2 Particle S-matrix}

\vspace{-1mm}

A useful perspective on the general class  \eqref{imscft}  of integrable deformations of a 2D CFT  is to view them as a limit of integrable deformations of a massive 2D QFT.
Massive 2D QFTs  are uniquely characterized by the spectrum of stable particles and their S-matrix. The presence of an infinite set of higher spin charges $P_s$ implies that the S-matrix factorizes into 2-particle S-matrices $S_{ab}(\theta)$, which depend on the difference $\theta = \theta_a -\theta_b$ between the rapidities of particles $a$ and $b$. Any integrable 2D QFT admits an infinite parameter family of deformations, defined by multiplying each 2 particle S-matrix $S_{ab}(\theta)$ with a so-called CDD phase factor
\es{cdd}{
S_{ab}(\theta) & \to e^{i\delta_{ab}(\theta)} S_{ab}(\theta)\,.
}
The most general allowed phase takes the form of a sum  $\delta_{ab}(\theta) = \sum_{s \in \N} \alpha_s \sinh(s \spc \theta)$ of  integer spin contributions.
The deformation parameters $\alpha_s$ of the 2-particle S-matrix are in one-to-one correspondence with the deformation parameters $\mu_s$ of the CFT action \eqref{imscft}.
In case only the lowest spin deformation is turned on, the CDD phase factor simplifies to $\delta_{ab}(\theta) = -{\mu\ov 4}\spc  m_{a} m_b \sinh(\theta)$, with $m_a$ and $m_b$ the mass of each particle. CFTs have only massless left- and right-moving excitations. To take the CFT limit, we thus boost particle $a$ and $b$ in opposite directions to the speed of light, while sending $m_a$ and $m_b$ to zero. In this limit the phase reduces to the product of the light-cone momenta of the two particles
\es{highlim}{
\delta_{ab}(\theta) & = \spc -\frac{\mu}8\spc m_a m_b \spc  e^{\theta_a - \theta_b} \spc = \spc -{\mu\ov 4}\spc p^{+}_a p^{-}_b \,.
}
In the unperturbed CFT left- and right-moving excitations pass through each other. The 2-particle S-matrix of the $T\bar T$ deformed CFT thus takes the simple form
\es{sab}{
S_{ab} & = e^{-i\mu\spc p^{+}_a  p^{-}_b/4 }\,.
}

This scattering phase \eqref{sab} is a toy version of the forward scattering amplitude of two highly boosted particles in 3+1-D Einstein gravity, as first studied by 't Hooft \cite{gravscat}. It describes the effect of a gravitational shockwave caused by the stress-energy of one particle on the trajectory of the other particle. To make this physical  interpretation explicit, consider a localized right-moving excitation $A(x^+)$ with small light-cone momentum $p_+$ and a left-moving mode $B_{p_-}$  in a momentum eigenstate with large light-cone momentum $p_-$. The 2-particle S-matrix \eqref{sab} expresses the property that the operators $A$ and $B$ do not commute, as they would in the undeformed CFT, but satisfy a non-trivial exchange relation of the form
\es{sabshock}{
A(x^+) \spc B_{p_-}\spc &  = \spc  B_{p_-} \spc A\le(x^+ - {\mu \spc p_-\ov 4}\ri)\,.
}
This exchange relation exhibits the effect of a gravitational shockwave created by the energetic left-moving mode $B$ on the position of the right-moving mode $A$.

Introducing the scattering phase \eqref{sab} has many interesting consequences.  We briefly mention two of these.

{\it Ground state energy.} Via the thermodynamic Bethe ansatz, one can
derive the ground state energy on a spatial circle with period $L$ as a function of $\mu$. One finds that \cite{nyugroup}
\es{gse}{
 E_0  ={2 L\ov \mu}\le(1-\sqrt{1+{ \pi c  \spc \mu \ov 6 L^2}}\spc \ri)\,. 
}
Note that this formula has a square root singularity and becomes imaginary for $-{\pi c \mu \ov 6} > {L^2}$.\footnote{See \cite{Mussardo:1999aj} for early work exploring this singularity in the thermodynamic Bethe ansatz.} Since we may interpret the period of the circle as an inverse temperature $L$, this indicates that for $\mu<0$, the theory breaks down above a critical temperature $\TT_{\!\spc \stars} = \sqrt{{- 6 \ov \pi c \mu}}$. This breakdown is called the Hagedorn transition in \cite{italians,nyugroup}. As we will see shortly, for the connection with holography we are interested in the opposite regime $\mu >0$. 

{\it Lyapunov behavior.} Equation \eqref{sabshock} is similar to the shockwave interaction that gives rise to the 
chaotic dynamics of black hole horizons. It seems plausible, therefore, that the $T \bar T$ perturbation can be used to make the Lyapunov growth of out of time ordered (OTO) correlation functions \cite{Roberts:2014isa,Maldacena:2015waa} at finite temperature more manifest.

\subsection{Energy spectrum}\label{sec:spectrum}
\vspace{-1mm}

The generalization of the exact result \eqref{gse} to arbitrary energy eigenstates, quoted in the introduction, can easily be derived from the Zamolodchikov relation \eqref{zamttb} as follows.
Consider the deformed CFT on a spatial circle parametrized by a angular coordinate $\theta$ with period $L$. 
Let $\ket{n}$ denote an energy and momentum eigenstate in CFT. Its energy $E_n$ and momentum $P_n$ take the general form
\es{erescale}{
E_n& = {{\cal E}_n\bigl(\mu/L^2\bigr)\ov L}\, , 
\qquad \qquad\  \ P_n = { 2\pi J_n\ov L}\, , \ \qquad \  \, J_n \in \Z\, .
}
In the CFT limit, we have $E_n^\text{(CFT)} L =  2\pi\bigl(\Delta_n + \bar\Delta_n - {c \ov 12}\bigr)$ and $J_n =  \Delta_n - \bar\Delta_n\in\Z$.  Since energy and momentum eigenstates are stationary and translation  invariant, the Zamolodchikov relation \eqref{zamttb} applies. It is useful to rewrite it as
\es{Key}{
\bra{n} T\bar{T}\ket{n}&=\bra{n} T\ket{n}\bra{n}\bar{T}\ket{n}-\bra{n} \Theta\ket{n}\bra{n}\Theta \ket{n}\\[1.5mm]
&=-\frac14\Bigl(\bra{n} T_{\tau\tau}\ket{n}\bra{n}T_{xx}\ket{n}-\bra{n} T_{\tau x}\ket{n}\bra{n}T_{\tau x} \ket{n}\Bigr)\,,
}
where we used the definitions and simple algebra. The stress tensor components have physical meaning as the energy density, pressure and momentum density. Hence we can express the right hand side in terms of physical properties of the spectrum:
\es{physical}{
\bra{n} T_{\tau\tau}\ket{n}&={E_n \ov L}\,,\qquad
\bra{n} T_{xx}\ket{n} = {\p E_n \ov \p  L}\,,\qquad
\bra{n}T_{\tau x} \ket{n}={iP_n \ov  L}\,.
}
The $i$ in the last formula follows from $T_{\tau x}=i(T-\bar{T})$.
The left-hand side of \eqref{Key} represents the $\mu$ dependence of the energy $E_n$, via the relation\footnote{To see this, note that in euclidean signature $H_\text{int} =  \int d\theta\, {\cal L}_\text{int}$ . This gives ${\p  \ov \p \mu}  \< H\> = \int \! d\theta \, \< \spc T \bar T\spc\> = L\, \< \spc T \bar T\spc\>.$}
\es{LHS}{
{\p E_n \ov \p \mu}\, &=  L \, \bra{n} T\bar{T}\ket{n}\,.
}
The relation \eqref{Key} thus combines into the following differential equation for $E_n$
\es{Burgers}{
0&= 4  {\p E_n\ov \p \mu} +{E_n } {\p E_n \ov \p L}+{P^2_n\ov L}\, .
}
As remarked in \cite{smirzam,italians}, this equation is formally identical to the forced inviscid Burgers equation.   Given that $E_n$ and $P_n$ have the form \eqref{erescale} 
and using the CFT value as initial condition, it is not hard to check that the solution to \eqref{Burgers} is given by 
\es{BurgersSol}{ 
 &\ \, E_n(\mu,L) L \, \equiv\, {\cal E}(\tilde{\mu})\, =\, {2\pi \ov \tilde\mu}\spc \le[\spc 1-\sqrt{1\spc -\spc 2 {\tilde\mu } M_n +\spc  {\tilde\mu^2}  J_n^2}\;  \ri] 
\\[2mm]
&M_n  =\De_n\! \spc +\bar\De_n\! \spc -{c\ov 12}\,, \qquad \ J_n =  \Delta_n\!\spc -\bar\Delta_n\, ,  \qquad \tilde \mu\equiv {\pi \mu\ov L^2}\,.
}
This relation reduces to the usual CFT value at $\tilde\mu \to 0$, and to the formula \eqref{gse} for $\Delta_n= \bar\Delta_n =0$.

\begin{figure}[t]
\begin{center}
\includegraphics[scale=0.83]{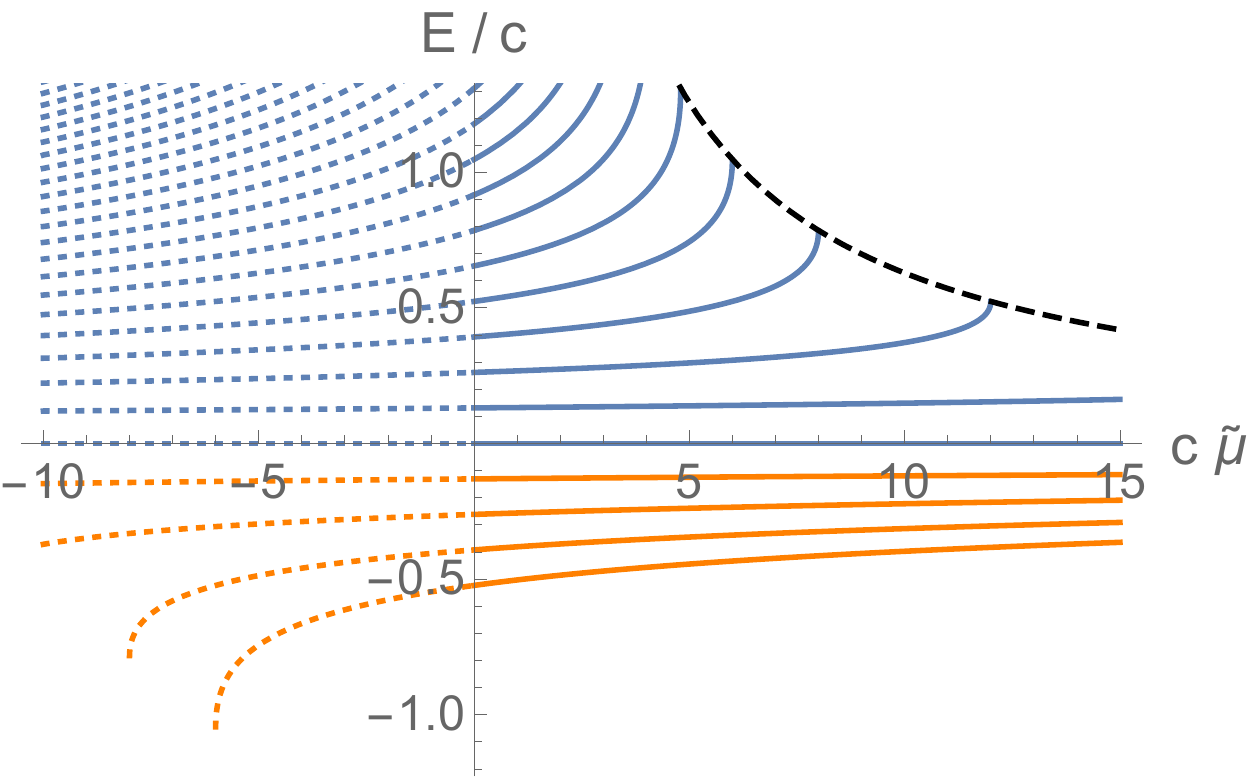}
\caption{The energy levels $E_n$ at $L=2\pi$ and $J=0$ as a function of $\mu$ for different values of $E(0) =  \De_n+\bar\De_n-{c\ov 12}$. States with $E(0)>0$ that correspond to black holes in holographic CFTs are plotted in blue, while low-lying states are plotted in orange. For $\mu>0$ that is the relevant regime in our study we used solid lines, while for $\mu<0$ the spectrum is plotted with dotted lines. The levels exhibit a square root singularity at the critical  value $\mu E(0) = 2\pi$. This indicates that, for given $\mu$, the energy spectrum of the deformed CFT is bounded by $E<  {8 \ov \mu}$, indicated on the plot by a dashed black line. \label{fig:spectrum}}
\vspace{-2mm}
\end{center}
\end{figure}

The spectrum as a function of $\tilde \mu$ takes the form plotted in Fig.~\ref{fig:spectrum} for $\De_n=\bar \De_n$. With an eye towards large $c$ CFTs, we have scaled the energies and $\tilde \mu$ by $c$. The lowest energy level plotted is the ground state. 
The state with $\De_n+\bar\De_n = {c\ov 12}$ (corresponding to the $M=0$ BTZ black hole) has zero energy independent of $\mu$.

\subsection{Thermodynamics}\label{QFTthermodynamics}

The formula \eqref{BurgersSol} has a nice scaling form and does not depend on the UV cutoff.
It can be read as describing the $\mu$ dependence of an energy level $E_n$ at fixed $L$, or as the variation of the energy  under an adiabatic
change in the circumference $L$ at fixed $\mu$. Note that $E_n(\mu, L, \Delta_n,\bar{\Delta}_n)$ is a monotonic function of $\Delta_n$ and $\bar{\Delta}_n$, so energy levels indeed do no cross as we vary $\mu$ or $L$. Hence the entropy remains $\mu$ independent and at high energy is given by the Cardy formula 
$S  =2\pi \sqrt{\frac c 6 (\Delta_n\! - {c \ov 24}) \spc} +\spc 2\pi \sqrt{\frac c 6 (\bar\Delta_n \! - {c \ov 24})}$ \cite{Cardy:1986ie}.

Equation \eqref{BurgersSol} exhibits a square root singularity for some critical value of $\tilde{\mu}$, called the `shock singularity' in \cite{smirzam}.\footnote{In the application of the Burgers equation to fluid mechanics, 
$E(\mu, L)$ represents the fluid velocity, with $\mu =$ time  and $L = $ position. The square root singularity then corresponds to the formation of a shock wave.} When $\Delta_n + \bar\Delta_n > {c \ov 12}$, the singularity occurs for positive value of $\mu$.
Its appearance  indicates the presence of a high energy cutoff, and seems to suggest that for given $\tilde \mu$ the spectrum of the $T \bar T$ QFT truncates above a critical value for the conformal dimension. The scale at which the spectrum truncates is where a naive analysis would have predicted locality to break down.\footnote{ The  dimensionless coupling constant, that indicates the scale at which the theory may break down, is the product of $\mu$ and the energy density, ${\tilde\mu } M$. This becomes $O(1)$ where \eqref{BurgersSol} has the square root singularity.}
Hence the deformed CFT on a cylinder has only a finite number of quantum states.
Since the Cardy entropy monotonically grows with energy,  we can also interpret this truncation as a bound on the total entropy of the system. The energy and entropy bound take the  form
\es{smax}{
E < E_\text{max} \spc =\spc {2 L \ov \mu}\,, \qquad \qquad S \, < \, S_\text{max} = L \sqrt{c \ov 6\pi \mu}\, .
 }
We will see that, on the gravity side, the state with maximal energy and entropy that saturates this bound corresponds to a maximal size black hole, that still fits inside the cutoff AdS space-time.

From now on we specialize to the non-rotating case $J=0$. The equation of state of the deformed CFT defines a relation between the energy $E$, the circumference $L$,  and the entropy $S$
\es{eostate}{ 
EL  - {\mu \ov 4} E^2 & = {3\spc S^2 \ov 2 \pi \spc c } \, . 
}
For $\mu=0$, this reduces to the usual Cardy formula.
From \eqref{eostate} we can derive other thermodynamic quantities, such as temperature and pressure, via the first law
\es{firstlaw}{
d E &\spc =\spc T dS \spc - \spc p dL \,
}
where the derivatives are taken while keeping the dimensionful coupling constant ${\mu}$ fixed.
We thus obtain the following relations between the entropy density
$s = {S/L}$, energy density $\rho= {E/L}$, pressure $p$, and temperature {\small $T$}
\es{temptress}{
{\rho/ \rho_{\spc \stars}} \, & = \, 1 - \sqrt{1 - {s^2/ s_\stars^2}}\,, \qquad \qquad\ \ \ \; p \,   = \, {\rho \ov {1 - \spc \rho/\rho_{\spc \stars} }}\,, \\[2mm]
p/\rho_{\spc \stars} & =  \spc \sqrt{1+ \TT^2/\TT_{\!\spc\stars}^2} \, -\spc 1 \,,  \qquad \quad {\TT/ \TT_{\! \spc \stars} }\, = \, {s/s_\stars \ov   {\sqrt{1 - s^2/s^2_{\spc \stars}}}}\, .
}
Here we introduced the critical values 
\es{critical}{
\rho_\stars & = {2 \ov \mu}\,, \qquad \qquad s_\stars = \sqrt{ 2\pi c \ov 3 \mu}\,,  \qquad \qquad T_\stars = \spc {\rho_\stars \ov s_\stars} = \sqrt{6\ov \pi c \mu}\,.
}
Note that all these critical values diverge for $\mu\to 0$. One easily verifies that  \eqref{temptress} reduces to the standard CFT relations in  this limit.
From equation \eqref{temptress} we verify the standard relation  $p  =- \rho + s \spc T$,
and obtain the free energy as a function of the temperature \cite{nyugroup}
\es{morethermo}{
F \spc = E - TS = \spc {2L \ov \mu}\le( \spc 1 \spc - \spc \sqrt{1 + \spc {\TT^2}/\TT_{\!\spc \stars}^2}\, \ri)\,.
}

The propagation speed $v_s$ of sound waves will play a central role in the comparison between the deformed CFT and gravity. For the non-rotating case $J=0$, we can compute $v_s$ via
\es{speedofsound}{
v_s &\spc =\spc \sqrt{\spc {\p p \ov \p \rho}}=\spc { {1}\ov 1-\spc \rho/\rho_{\spc \stars}} \spc =\spc  {1 \ov \sqrt{1\spc -\spc 2 {\tilde\mu } M }}\spc \,.
}
with $\tilde{\mu} = {\pi \mu\ov L^2}$.  
We observe that for $\mu>0$ sound waves propagate at superluminal speeds, and moreover that the temperature, pressure, and the sound speed all diverge at a critical values for the energy and entropy density. Near this critical value, the compressibility and the heat capacity of the system both go to zero.
This singular behavior is another indication that the deformed CFT has a UV cutoff.
As we will see in the following sections, the superluminal sound speed and the divergence of pressure and temperature  all have a direct physical interpretation in the gravity dual description.

\subsection{Equivalence to Nambu-Goto}

There exists an instructive relationship between the deformed CFT  and the Nambu-Goto (NG) string. This relationship is most explicitly understood for the case that the CFT has central charge $c=24$, where it can be shown to be a direct equivalence with the worldsheet theory of critical string theory. Moreover, this reformulation makes manifest that, in this special case, the $T \bar T$ deformed CFT represents a well defined, unitary and exactly soluble quantum system. This observation could help alleviate some possible worries the reader may have about the UV completeness of the theory.

Starting with some general CFT with $c=24$, we define the deformed theory by adding two free massless scalar fields $X^+$ and $X^-$. The total action reads
\es{xpmdef}{
S_\text{QFT} &\, = \, S_\text{CFT}\spc + \spc {1\ov 2\mu} \int \! d^2x \,\spc \p_\alpha X^+\spc \p^\alpha X^-\,.
}
In the analogy with a string worldsheet theory, the free fields play the role of light-cone target space coordinates, whereas the CFT represents some general (abstract) 24-dimensional target space. Note that the kinetic term of the scalars $X^\pm$ has the opposite sign to the usual NG string. Just like one would in the NG formulation of string theory, we supplement the free field equation of motion $\p_u \p_v X^\pm = 0$ with the Virasoro conditions\footnote{Here we use the terminology Nambu-Goto CFT somewhat loosely. In case the $c=24$ CFT is described by 24 massless free scalars, the action \eqref{xpmdef} combined with the Virasoro constraints  \eqref{ngeom} yield, via the usual Goddard-Goldstone-Rebbi-Thorn treatment, the standard Nambu-Goto action for a critical string in $D=26$ dimensions. We can consider a more general case, however, in which 24 of the 26 target space dimensions are replaced by a general target space, described by a general CFT with central charge $c=24$. This is still a critical string theory, with a well defined world sheet theory. We call the world sheet theory of this more general string theory, defined by \eqref{xpmdef} and \eqref{ngeom}, a Nambu-Goto CFT. }
\es{ngeom}{
-\p_u X^+ \p_u X^-   + \mu   \spc T^\text{\,CFT}_{uu} & = 0\, , \qquad  \  -\p_v X^+ \p_vX^-   + \mu  \spc T^\text{\,CFT}_{vv} = 0\, .
}
 Here $u$ and $v$ denote the light-cone coordinates on the worldsheet. The constraints \eqref{ngeom}  implement gauge invariance under arbitrary conformal transformations $(u,v) \to (\tilde u(u),\tilde{v}(v))$. We can use this invariance to choose special worldsheet coordinate $(x^+,x^-)$ such that\footnote{We thank Juan Maldacena for a helpful discussion on the usefulness of this particular light-cone gauge.}
\es{lcgauge}{
\p_+ X^+ & = \, \p_- X^- = \, 1\, .
}
This gauge choice is analogous to the light-cone gauge in string theory. It identifies the worldsheet light-cone coordinates with the respective chiral halves of the target space light-cone coordinate fields:  $X^+(x^+,x^-) = x^+ + \tilde{X}^+(x^-)$ and $X^-(x^+,x^-) = x^- + \tilde{X}^-(x^+)$. The other chiral halves of the light-cone fields are determined by integrating the Virasoro conditions 
\es{virconst}{
-\p_-  {X}^+ & \! + \, {\mu} \, T^\text{\,CFT}_{--} = 0\,,  \qquad \ \ \ \qquad \ \ X^+ \!  = \, x^++   {\mu} \int^{x^-}\!\!\! T^\text{\,CFT}_{--}\,,\\[-3.25mm] &\  \qquad\qquad\qquad \quad\quad \implies \\[-3.25mm]
 -\p_+ {X}^-   & \! + \, {\mu}\, T^\text{\,CFT}_{++} = 0\, ,
\,  \qquad\ \ \  \qquad \ \ X^- \! =\,x^- + {\mu} \int^{x^+}\!\!\! T^\text{\,CFT}_{++}
\, . 
}
Following the GGRT treatment of the NG string \cite{GGRT}, equations \eqref{virconst} provide the quantum definition of the light-cone coordinate fields.
The self-consistency of this identification at the full quantum level is well established for $c=24$, and forms the basis of the no ghost theorem for the critical NG string \cite{brower,scherk}. For our context, this theorem provides a direct proof that the $T \bar T$ deformed CFT with $c=24$ is a well defined unitary quantum theory, in which all Hilbert states have positive norm.

The above reformulation provides an exact non-perturbative definition of the $T\bar T$ deformation of a CFT with $c=24$.  There are several ways to see that the two systems are indeed equivalent. 
A first simple check is that the leading order interaction term in the NG action indeed takes the form of a $T \bar T$ interaction \cite{nyugroup,italians}. 
Conversely, if we insert equations \eqref{lcgauge} and \eqref{virconst} into \eqref{xpmdef}, it becomes equal to the action \eqref{ttbardeformMink}. Note that the mapping between the two actions involves a coordinate transformation from the worldsheet coordinates $(u,v)$ to dynamical target space coordinates $(X^+,X^-)$ that back react to stress-energy. From equation \eqref{virconst} it is also clear why the above coupling of the CFT to the dynamical light-cone coordinates leads to non-trivial scattering phase 
and exchange relations between left and right-movers of the form \eqref{sab} and \eqref{sabshock} \cite{svv}. Another direct check is that the energy levels in Nambu-Goto theory are precisely of the form \eqref{BurgersSol}.\footnote{As we will see, however, the sign of $\mu$ required for our holographic interpretation is {\it opposite} to the standard sign in the Nambu-Goto string. This seems 
puzzling in view of the result of \cite{adams} that in non-linear QFTs  with NG or DBI actions, only one sign of the higher derivative coupling leads to a consistent, causal theory. Our proposed resolution is that the notion of causality depends on how one defines the detectors that measure signals. 
There are two types:  a) detectors anchored to the fixed  coordinates $(u,v)$, or
b) detectors attached to the dynamical coordinates $(X^+, X^-)$. Standard NG theory is causal with respect to detectors of type a), while the holographic theory is causal for detectors of type b). We'll return to this point in the next sections.}

\section{Gravitational Energy and Thermodynamics}\label{thermodynamics}
\vspace{-1mm}

In this section we will make a comparison between the thermodynamic quantities of the $T \bar T$ deformed CFT and those of a BTZ black hole in a region of AdS with finite radial cutoff $r=\rrc$. We will summarize the derivation of the total quasi-local gravitational energy of the black hole as a function $\rrc$ \cite{brownmann}, and show that it agrees with the QFT result \eqref{BurgersSol}. The metric of a BTZ black hole with mass $M$ and angular moment $J$ (with left- and right inverse temperature $\beta_\pm$) can be written as
\es{BTZ}{
ds^2  & = - f^2(r) dt^2+f^{-2}(r) dr^2+r^2\le(d\theta- \omega(r)\,dt\ri)^2\\[2mm]
&\!\!\!\!\!\! f^2(r)    ={ r^2 } - 8G M + {16 G^2 J^2 \ov  r^2}\, , \ \qquad 
\omega(r) = {4G J \ov r^2}\, ,\\[2mm]
M &={r_+^2+r_-^2\ov 8G}\,, \quad\  \ J={r_+ r_- \ov 4G}\,, \quad\ \   \ \beta_\pm={ 2\pi \ov  r_+ \mp r_-}\,.
}

The gravitational action for a 3D space-time with negative cosmological constant $\Lambda = -1$ and with a time-like boundary $B$  is
\es{gaction}{
S & = {1 \ov 16\pi \GN} \int\! d^3 x\, \sqrt{-g_3}\spc \bigl( R + 2\bigr) - { 1\ov 8\pi \GN} \int_B \! d^2 x \, \sqrt{-g} \spc (K + 1)\,,
}
where $K$ denotes the extrinsic curvature of the boundary. For simplicity, we omit all contributions due to matter fields:
we assume that the classical matter sources can all be self-consistently turned off. The classical solutions of \eqref{gaction} are therefore locally pure $AdS_3$. We are interested in all solutions in the neighborhood of a general rotating BTZ black hole solution with a static boundary $B =  \{ r=\rrc\}$. We impose Dirichlet boundary conditions, by fixing the form of the boundary metric 
\es{bmetric}{
ds^2\bigl\vert_{B} & = g_{\alpha\beta} dx^\alpha dx^\beta =  - N^2 dt^2 + e^{2\varphi}(d \theta - \omega dt)^2\,.
}
By solving the bulk equation of motion with given boundary condition, the action \eqref{gaction} becomes a functional $S[g]$ of the boundary metric.
Note that in \eqref{gaction} we have included a boundary cosmological constant, tuned such that it cancels the leading volume divergence of the bulk action.

The gravitational energy of the black hole space-time is defined in terms of the variation of the action functional $S[g]$. In the parametrization \eqref{bmetric}, this variation takes the general form \cite{brownmann}
\es{deltas}{
\delta S &  = \int_B\! d^2x\, \pi^{\alpha\beta} \delta g_{\alpha\beta} = \int_B \! d^2 x\, \sqrt{-g}\spc \bigl( - \epsilon \delta N - j \delta \omega +  p \spc \delta \varphi\bigr)\,.
}
The quantities $\epsilon$, $j$ and $p$ can respectively be interpreted as gravitational energy density, momentum density, and pressure, as measured on the boundary $B$.  The total energy is thus obtained by 
integrating the energy density over the spatial section of $B$
\es{edefint}{
E  & = \oint\! d \theta \, e^{\varphi} \epsilon\,.
}

We now want to apply this formalism to compute the quasi-local energy of a rotating BTZ black hole \eqref{BTZ} with mass $M$ and angular momentum $J$ inside a Dirichlet wall at $r_c$.
 Following \cite{brownmann} we get
\es{BTZcutoff}{
E&={r_c \ov 4 \GN}\le[1-\sqrt{1-{8\GN M\ov \rrct}+ { 16 \GN^2 J^2\ov \rrc^4}}\, \ri]\,.
}
In the limit $\rrc \to \infty$, this formula reduces to $E =  M$.
If we multiply \eqref{BTZcutoff} by the circumference of the circle $2\pi\rrc$, we obtain a formula for the dimensionless quantity ${\cal E} =2\pi \rrc \spc E $, which perfectly matches with the result~\eqref{BurgersSol} obtained on the QFT side,
provided we identify
\es{Match}{
{M }&=M_n=\De_n+\bar\De_n-{c\ov 12}\,, \qquad \quad {J} =J_n=\De_n- \bar\De_n\,, \qquad  \quad \tilde\mu ={4 \GN\ov \rrct}={6\ov c}\, {1\ov \rrct}\,.
}
The first two identifications are completely standard in AdS/CFT. We again recover our proposed identification between the deformation parameter $\tilde\mu$ the cutoff radius $\rrc$ (see footnote \ref{foot2}).

Note that the definition of the quasi-local energy only makes use of the intrinsic geometric properties of the cutoff space time and its boundary metric, and does not make any reference to the coordinate system of the asymptotic AdS observer. The thermodynamic quantities do not change under the change of coordinates in \eqref{bmetric}, $t'=a t,\, \theta'=\theta+ b \spc t$, which respects the periodicity $\theta\sim\theta+2\pi$. In other words, the formula \eqref{BTZcutoff} represents the total energy as measured in the canonical time coordinate of an observer living on the boundary $B$, in which the boundary metric has the form \eqref{bmetric} with lapse $N=1$. This is the right definition for comparison with the definition of energy in the deformed CFT. As explained in the previous section,  due to the combined effect of the $T \bar T$ interaction and turning on a finite temperature, the metric (as defined as in footnote \ref{foot1}) gets renormalized relative to the metric of the undeformed CFT. 
The total energy of the QFT is defined as the integral of $\< T_{00}\> = {2i \ov \sqrt{g}} {\delta  \ov {\delta g^{00}}}\log Z_\text{QFT}$, defined using the renormalized metric.
This is the quantity that matches between both sides. Combined with the known correspondence between the Bekenstein-Hawking entropy
\es{bhent}{
S = {\pi r_+ \ov 2 G}
}
 and the Cardy entropy of the CFT, 
this match between the energy spectra establishes a complete correspondence between the thermodynamical properties of  the deformed CFT and the BTZ black hole with a sharp radial cutoff. 

The holographic duality gives a new physical perspective on  the square root singularity  \eqref{energylevel} in the energy levels of the $T \bar T$ deformed CFT.  On the AdS side, this singularity in the gravitational energy \eqref{BTZcutoff}  occurs when the Dirichlet wall approaches the event horizon of the BTZ black hole. For given $\mu$ and $\rrc$, the critical behavior indicates an upper bound on the total energy and entropy
\es{esbound}{
E < E_{max} = { \rrc \ov 4\GN}\,, \qquad \qquad
S \, < \, S_\text{max} =  {\pi \rrc \ov 2 \GN}\, ,
 }
which are saturated in the limit that the BTZ black hole completely fills out the space-time inside the wall at $r = \rrc$.
While $E$ and $S$ both remain finite in the limit, the temperature $T$ and pressure $p$ both diverge at this critical value.
To compute both quantities, let us rewrite the QFT equation of state \eqref{eostate}  in bulk notation
\es{geostate}{ 
E \rrc  - {2 \GN} E^2 \spc & =  \spc {\spc \GN S^2 \ov \spc 2 \pi^2 }  \,.
}
Using the first law \eqref{firstlaw}, and the fact that the length of the Dirichlet wall is $2\pi \rrc$, we derive that
\es{thawking}{
T  & =\spc {r_+\ov 2\pi\spc (\rrc -\spc 4\GN E)} \spc = \spc {r_+ \ov 2\pi \sqrt{\rrct - r_+^2}}\,, \qquad \\[2mm]
 p   & =   \spc {E\ov 2\pi (\rrc  -\spc 4\GN E)}  = \spc  {E \ov 2\pi \sqrt{\rrct -r_+^2}}\,.
}
The divergence in the temperature is explained by the usual Tolman relation, or equivalently, by the fact that a static observer at the boundary must in fact undergo a uniform acceleration equal to $a = {r_+ / \sqrt{g_{00}} }$, with $g_{00} = {\rrct - r_+^2}$, in order to stay at constant radius $r=\rrc$. This acceleration diverges at the horizon, and via the Unruh effect, the static observer thus sees a black hole atmosphere as an incompressible fluid with a diverging temperature and pressure at the horizon. 

From second equation in \eqref{thawking} we directly compute the sound speed via
\es{gravsound}{
v_s = \sqrt{\p p \ov \p \rho} = {1 \ov {1 - {4 \GN E/ r_c}}} = {1 \ov \sqrt{1- r_+^2/r_c^2}}\,,
}
which should be compared with the QFT result \eqref{speedofsound} obtained earlier. We now see more directly  that the divergent sound propagation speed for $r_c = r_+$ arises from the incompressibility of the near horizon black hole atmosphere. We will study the propagation speed more closely in the next section.

\section{Signal Propagation Speed}\label{sec:speed}
\vspace{-1mm}
In this section we review Cardy's argument \cite{cardy}  that  in the $T\bar{T}$ deformed CFT at finite temperature signals propagate faster or slower than the speed of light. We then consider the rotating state, and in the Appendix the case in which the expectation values of the stress tensor are arbitrary functions of the light-cone coordinate.
We then compare with the prediction from our proposed holographic dictionary, and find a precise match. 

\subsection{Propagation speed from QFT}\label{speed-hs}
To linear order in the deformation parameter $\mu$, one can regard the deformed CFT \eqref{ttbardeform} as  an undeformed CFT coupled to a Gaussian random background metric. 
The metric fluctuations represent two spin 2 Hubbard-Stratonovich (HS) fields $\fxi, \bar \fxi$:
\es{Cardy}{
S_\text{QFT} & \spc =\spc S_\text{CFT}+\int d^2x \ \le[-{\fxi \bar \fxi \ov\mu}+ \bar\fxi \spc T+\fxi\spc \bar T\ri]\,.
}
This HS representation can be extended to finite values of the deformation parameter $\mu$ via the more exact formula  \eqref{legendre} quoted in the introduction. In this section we restrict ourselves to the linearized regime of small~$\mu$.
The action \eqref{Cardy} describes the CFT in the random metric 
$ds^2= dz d\bar{z} +{\bar\fxi\ov 2} dz^2+ {\fxi\ov 2} d\bar{z}^2$, or after
analytic continuation to Lorentzian signature via $z\to x^-,\, \bar{z}\to -x^+$,
\es{randomMetric}{
ds^2= -dx^+ dx^-+ {\fxi\ov 2} (dx^+)^2 +{\bar\fxi\ov 2} (dx^-)^2\, .
}

For later reference, note that to linearized order we may write this metric as $ds^2 = g_{\alpha\beta} dx^\alpha dx^\beta$~with
\es{randmetric}{
& g_{\alpha\beta} \spc =\spc
 \eta_{ab} v^a_\alpha v^b_\beta\,,\qquad \qquad
  v^\pm_\alpha  \spc \equiv  \, e^\pm_\alpha  + f^\pm_\alpha \,,
}
where $e^\pm_\alpha dx^\alpha\! = dx^\pm$ specifies the background metric and  $f^+_\alpha dx^\alpha\! =- {1\ov 2} \bar{f} dx^-$ and $f^-_\alpha dx^\alpha \! =- {1\ov 2}  f dx^+$ denotes the Gaussian fluctuation. The notation \eqref{randmetric} will be useful later on, as it will allow us to vary the background metric $g_{\alpha\beta}  = \eta_{ab} e^a_\alpha e^b_\beta$ independently from the fluctuating metric \eqref{randmetric}. Indeed, it will be important to keep track of the distinction between the two metrics and their corresponding light cones.
The null geodesics of the fluctuating metric are specified~by
\es{lcones}{
v^+_\alpha dx^\alpha \spc & \equiv \spc dx^+ - { f\ov 2} dx^- = 0 \,, \qquad {\rm or} \qquad 
v^-_\alpha dx^\alpha \spc \equiv \spc  dx^- - {\bar f\ov 2} dx^+ = 0\, .
}

 Let us consider a class of quantum states with non-zero expectation value of the stress tensor. 
Examples of such states are a thermal density matrix with finite inverse temperature $\beta$, or a semi-classical  coherent state, in which the left- and right-moving component of the stress tensor have some general position dependent expectation value $\<\spc T_{++}(x^+)\spc \>$ and $\< \spc {T}_{--}(x^-) \spc \>$. In the above HS representation of the deformed CFT, the fluctuating fields $\fxi,\bar \fxi$ attain the saddle point value\footnote{Note that by going to Lorentzian signature we introduce a sign, $\<\spc T\spc \> =-\<\spc T_{--}\spc \>\,,\,  \<\spc \bar T\spc \> =-\<\spc T_{++}\spc \>\,$.}  
\es{ExpVal}{
\<\spc \fxi \spc \>&=-\mu \spc \<\spc T_{++}\spc \> \,, \qquad \<\spc \bar \fxi\spc \>=-\mu\spc \<\spc T_{--}\spc \>\,.
} 
Again we see that the presence of stress-energy affects the effective space-time geometry of the deformed CFT. This leads to a renormalization of the 
propagation speed of the left- and right-moving
degrees of freedom.  We can compute this effect using the null geodesics \eqref{lcones}.
At the stationary point of the random metric \eqref{randomMetric}, the propagation speeds are renormalized to
\es{propspeed}{
	v_+& \simeq 1+ \mu\corr{T_{++}}, \qquad \qquad v_- \simeq 1+ \mu\corr{T_{--}}\,.
}
This is the result that we argued for in the introduction \eqref{newLC}.
For the special case of a thermal state with left and right-moving inverse temperature $\beta_\pm$ we have (c.f. \cite{cardy})
\es{thermospeed}{
\<\spc T_{\pm\pm}\> & =   {\pi\, c\ov 12 \beta_\pm^2}\,   \qquad 
\, \implies \qquad v_\pm  \spc \simeq \spc 1\spc + \spc {\pi c \spc \mu \ov 12 \beta_\pm^2}\, .
}

For $\mu > 0$, equation \eqref{thermospeed} and \eqref{speedofsound} represent superluminal speeds, while for $\mu<0$ the renormalized speed is subluminal. The microscopic explanation of this effect is as follows \cite{cardy}. The $T \bar T$ term for $\mu<0$ leads to an attractive interaction and a positive time delay whenever a left- and right-moving particle collide. At finite temperature,
 the particles scatter off of a sea of quasi-particles, and the resulting time delay  reduces the propagation speed. For $\mu>0$, on the other hand, the repulsive inter-particle interaction leads to a time advance, and the accumulative effect of the scattering enhances the propagation speed. It is important to emphasize, however, that this speed is superluminal only relative to the fixed background metric, and that physics remains causal relative to the fluctuating metric \eqref{randomMetric}. The UV limit of the $T \bar T$ deformed theory with $\mu>0$ does not define a usual local CFT, but nonetheless behaves like a causal theory similar to 2D quantum gravity.

\subsection{Propagation speed from thermodynamics}

We can also compute the propagation speed using the thermodynamic equation of state. As we have seen in subsection \ref{QFTthermodynamics}, by combining the exact formula \eqref{BurgersSol} for the $\mu$ dependence of the energy eigenvalues with the Cardy  entropy formula of the CFT, we can derive the  exact $\mu$ dependence of all thermodynamic quantities,
including the speed of sound \eqref{speedofsound}. Here we generalize the discussion to the rotating case. 

In the rotating system, the equation of state \eqref{geostate} extends to a relation between the energy $E$, entropy $S$, radius $L$, and angular momentum $J$ of the form
\es{eostnew}{
{E } L -  {\mu \ov 4}  \le(E^2 - {4\pi^2 J^2 \ov L^2}\ri)  \, =\, {{\; 3\spc\spc S^2} \ov  2\pi c }\spc  \, +\,   \spc {\pi^2c\spc  J^2\ov {\, 3 \spc \spc S^2}}  \,.
}
The first law of thermodynamics generalizes to
\es{gfirstlaw}{
d E = T dS - p dL + \Omega dJ \,
}
where $\Omega$ is to be identified with the angular rotation speed, possibly up to a constant shift. Here the derivatives are taken with ${\mu}$ fixed.

Let us first consider the system in the large $L$ limit, while keeping the energy density $\rho = E/L$, (angular) momentum density $j = J/L$ and entropy density $s = S/L$ fixed. We can then drop the term proportional to $J^2/L^2$ in \eqref{eostnew}, since it becomes small relative to the other terms. In this sense, we can think of this term as a finite size correction.  The pressure and angular chemical potential $\Omega$ in the large $L$ limit  can be expressed as
\es{pzero}{
\qquad \bar{p} & \spc =\spc  {\rho   \ov 1 - \mu\spc\spc \rho/2}\,, 
\qquad\qquad \bar\Omega \spc  =\spc   {2\pi^2c\spc  j\ov {\, 3 \spc \spc s^2 ( {1- \mu \rho/2})}}  \,.
}
We notice that the relationship between $p$ and $\rho$ is identical to the non-rotating case. In particular, we deduce that the speed of sound in this limit is still equal to 
$\bar{v} = \sqrt{\p \bar{p} \ov \p \rho} = {1 \ov 1- \mu \rho/2}$.
This speed is the same in both directions. This seems surprising, since a  priori one would be inclined to interpret $\bar\Omega$ as the angular rotation speed of the QFT fluid. However, the corresponding linear speed $\bar{u}\spc  =  \spc {L \ov 2\pi}\, \bar{\Omega}$ diverges in the large $L$ limit. So we will instead interpret $\bar{\Omega}$ as an off-set that has to be subtracted from $\Omega$ in order to get the physical angular velocity. So we will apply the redefinition  $\Omega_\text{new}   =  \Omega_\text{old} - \bar{\Omega}$.
 
With this new definition, let us include the finite size term $J^2/L^2$. The pressure $p$ and angular velocity $\Omega$ at finite $L$ are given by
\es{pressure}{
\qquad p  &  =    {\rho -   {2\pi^2 \mu \, j^2/ L^2} \ov 1 - \mu\spc\spc \rho/2 } \,, \qquad \qquad \Omega  \spc = \spc  -  {2\pi^2 \mu \spc  j\ov {L^2 ( {1- \mu \rho/2})}}\,.
}
The first relation reduces to $p=\rho$ at $\mu=0$, as it should. 

We would like to extract the propagation speeds from the two formulas \eqref{pressure}. First we note that
\es{speedsquare}{ 
\le({\p \spc p \ov \p \rho}\ri)_L&\, =\, {1- \pi^2 \mu^2 j^2/L^2 \ov (1-\mu\rho/2)^2} \, =\,  {v}_+ {v}_-
}
with
\es{twospeeds}{
{v}_\pm & \, = \,{1\pm   \pi \mu \spc j/L \ov 1 - {\mu} \spc \rho/2\spc} \, =\, \spc {1\pm  \tilde \mu J \ov \sqrt{1\spc -\spc 2 {\tilde\mu } M +\spc \textstyle  {\tilde\mu^2}  J^2}  }\,,
}
where $\tilde{\mu}  = {\pi \mu \ov L^2}.$ Here in the second step we used equation \eqref{BurgersSol}. 

It is reasonable to interpret the quantities $v_\pm$ as the left- and right-moving signal propagation speeds in the rotating deformed CFT. 
This interpretation is supported by the fact  that the  angular rotation frequency $\Omega$ and the left- and right velocities $v_\pm$ are related via
\es{uomegarel}{
v_\pm  \spc & = \spc \bar{v}\spc \mp \spc {\Omega L \ov 2\pi}  
}
with $\bar{v} = 1/\sqrt{1\spc -\spc 2 {\tilde\mu } M +\spc \textstyle  {\tilde\mu^2}  J^2}$. 
As we will see shortly, the formula \eqref{twospeeds} for the propagation speeds agrees with the renormalized velocity computed via the gravity dual.

 \subsection{Propagation speed from gravity}\label{sec:gravityspeed}
 
The renormalization of the propagation speed has a direct interpretation in the dual gravity theory as the statement that metric perturbations of a BTZ black hole with Dirichlet boundary conditions at the fixed $r=r_c$ surface travel at superluminal speeds relative to the fixed boundary metric. 
The idea of the following calculation was introduced by Marolf and Rangamani in \cite{marolf}. 

Consider a BTZ black hole of mass $M$ and angular momentum $J$, with  the space-time metric is given in equation \eqref{BTZ}, surrounded by a Dirichlet wall at $r = \rrc$. Now consider a fluctuation in the location of the boundary surface of the form  $r_c \to r_c + \delta r(t,\theta)$. The Dirichlet boundary condition requires that the induced metric on the perturbed boundary surface remains flat.
Computing the Ricci scalar of the induced metric on the perturbed boundary surface and expanding to linear order in the perturbation, one deduces  that $\delta r(t,\theta)$ satisfies the linear wave equation
\es{propo}{
R\le(r_c + \delta r(t,\theta)\ri)=-{2\ov  r_c\spc f^2(r_c)}\le[-\p_t^2+ \p_\theta^2\ri]\delta r(t,\theta) &\spc =\spc 0\,. 
}
We see that, perhaps somewhat expectedly, the fluctuation $\delta r(t,\theta)$ describes a wave propagating along light-like trajectories $dt = \pm d\theta$, as measured in the coordinate system anchored to asymptotic infinity. These light-like trajectories are superluminal relative to the metric on the cutoff surface itself. To compute their speed as seen
by an observer on the cutoff surface $r=r_c$, let us introduce coordinates $t_c,\, \theta_c$, so that the induced metric on surface is proportional to the standard flat metric  $ds^2\vert_{r=r_c} =  -dt_c^2 + d\theta_c^2$. In the $t,\, \theta$ coordinates the  induced metric on the wall is
\es{InducedMetric}{
ds^2\vert_{r=r_c} =-f^2(r_c) dt^2+r_c^2\le(d\theta- \omega(r_c)\,dt\ri)^2\, =r_c^2\le( -dt_c^2 + d\theta_c^2\ri)\,.
}
The change of coordinates that preserves the $2\pi$ periodicity of $\theta$ is: 
\es{CoordChange}{
dt= {r_c \spc  \ov f(r_c)} \, dt_c\,, \qquad \qquad  d\theta=d\theta_c - \Omega(r_c) dt_c\,, \qquad  \quad \Omega(r_c) \spc \equiv\spc -{r_c \spc \omega(r_c) \ov f(r_c)}\,.
}
The quantity $\Omega(r_c)$ is the rotation speed due to the frame dragging effect of the rotating black hole  as experienced at the cutoff surface $r=r_c$. It should be compared with the thermodynamic quantity $\Omega$ given in equation \eqref{pressure}.
The propagation trajectories are
\es{supertraject}{
d t & = \pm d\theta  \qquad \implies \qquad {r_c \spc \ov f(r_c) } \, dt_c\spc =\spc \pm\le( d\theta_c - \Omega(r_c) dt_c \ri)\,.
}
We read off that the left- and right-moving parts of the wave $\delta r(t_c,\theta_c)$  propagate with velocity
\es{propvv}{
v_\pm &\spc =\spc {r_c \spc \ov f(r_c)} \spc \mp \spc \Omega(r_c) \, = \, 
 { { 1 \pm {4G J \ov r_c^2}}\ov \sqrt{1  - {8G M \ov r_c^2} + {16 G^2 J^2 \ov  r_c^4}}}\,.
}
This result precisely matches with the signal propagation speed \eqref{twospeeds} computed from the thermodynamics of the deformed QFT,  provided we identify 
$\tilde\mu = {4\GN\ov r_c^2}$, which via the relations $\tilde\mu = {\pi \mu \ov L^2} = {\mu \ov 4\pi}$ (setting $L=2\pi$) and $\GN = {3 \ov 2c}$ reproduces the identification $\mu = {24\pi \ov c}\spc{1\ov \rrct }$ announced in the introduction. Specializing to the $J=0$ case, equation \eqref{propvv} matches with \eqref{speedofsound}.  In the limit of large $r_c$, from \eqref{propvv} we obtain $
v_\pm  \simeq \, \spc 1\spc +\spc {2\pi^2 / \rrct \beta_\pm^2}\,,$
which coincides with the field theory result \eqref{thermospeed}.

We generalize the analysis of signal propagation speed to a more general class of states in which the expectation value of the stress tensor has some arbitrary position dependence. The computation is presented in the Appendix, and we get the following linearized result 
 \es{LCAdS2}{
 v_\pm& \simeq \, 1+{16\pi G \ov  r_c^2}\,  \bigl\< T_{\pm\pm}(x^\pm)\bigr\>\,.
  }
The agreement between this result and the propagation speed \eqref{propspeed} computed in the deformed CFT is evidence that our proposed holographic dictionary extends to localized stress-energy perturbations.

From the gravity side, it still seems somewhat unsettling that the fluctuations of the boundary surface propagate at speeds that appear to violate boundary causality.
So some clarifying comments may be in order. First we note that the speed \eqref{propvv} is equal to the inverse of the blackening factor, that relates the light-cone at $r=\rrc$ to light-cone at asymptotic infinity in AdS. In other words, the speed \eqref{propvv} coincides with the light propagation speed at the asymptotic AdS-boundary. Somehow, the cutoff AdS space-time inside the Dirichlet wall has memory of the asymptotic light-cone, even though the asymptotic region is no longer there. A partial explanation for this phenomenon is that the propagating fluctuation described by \eqref{propo} represents a boundary graviton mode. In spite of its name, a boundary graviton is not literally localized at the boundary of AdS, but instead represents a non-local geometric degree of freedom, encoded in the diffeomorphism that relates the uniformizing coordinate systems at the UV and IR boundaries of the AdS space-time. This diffeomorphism and the boundary graviton modes are topological excitations, in the sense that their propagation speed is insensitive to the introduction of the Dirichlet wall. Hence a bulk observer in AdS can not detect these graviton modes as superluminal localized excitations that violate local micro-causality constraints.

To gain further insight, it is instructive to view the propagation velocity from the perspective of information spreading in the QFT, as the speed by which a small perturbation in an equilibrium thermal state delocalizes throughout the system \cite{douglasmark}. Suppose we act with a light local operator ${\cal O}(x,t=0)$ on the thermal state. The strongly coupled QFT dynamics delocalizes the perturbation over a region $\Sigma(t)$ with radius $R(t)$ that grows linearly with time with the butterfly velocity $v_B$  \cite{Shenker:2013pqa,Roberts:2014isa}.  To determine $v_B$ we look for the smallest region that contains sufficient information to reconstruct ${\cal O}(x,0)$. In the gravity dual, we can evaluate $v_B$ via the holographic postulate that the QFT state inside the boundary subregion ${\Sigma}(t)$ completely describes the bulk subregion $B_\Sigma$ contained within the Ryu-Takayanagi (RT) \cite{Ryu:2006bv,Ryu:2006ef} minimal surface associated with $\Sigma(t)$. The thermal state after acting with ${\cal O}(x,0)$ is described by a BTZ black hole with a small particle that falls towards the horizon. The smallest boundary subregion $\Sigma(t)$ that contains the information created by ${\cal O}(x,0)$ after time t is simply the smallest region such that the corresponding RT surface still contains the bulk particle \cite{douglasmark}. 

\begin{figure}[t]
\begin{center}
\includegraphics[ height=4cm, width=12cm]{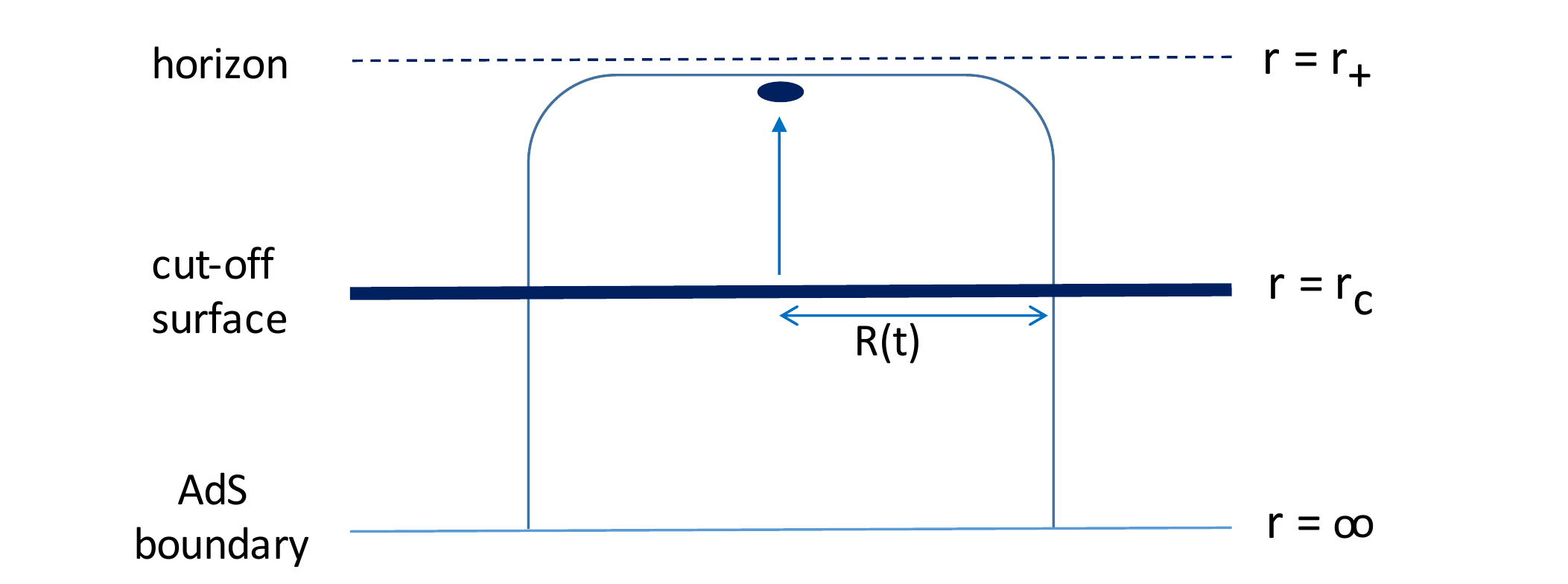}
\caption{As a localized wave approaches the horizon, the minimal RT surface that contains the excitation at time $t$ extends  along the horizon over a distance $R(t)$ that grows linearly in time. }\label{fig:wedge}
\vspace{-2mm}
\end{center}
\end{figure}

This situation is depicted in figure \ref{fig:wedge}. As time passes, the particle falls exponentially slowly towards the horizon. The minimal RT  surface that contains the particle at some late time $t$ must follow a path that stays exponentially close to the horizon over a distance $R(t)$, that is roughly equal to the size of the associated boundary subregion. The size of the boundary subregion grows linearly in time as  $R(t) \sim v_B t$. The butterfly velocity $v_B$ is equal to the speed of light as measured on the  asymptotic AdS boundary at $r=\infty$. This means that the subregion on the cutoff surface at $r=r_c$ grows with superluminal speed compared to the light speed measured at $r=\rrc$. This speed is equal to the signal propagation speed found in \eqref{propvv}.\footnote{This result comes from comparing the opening angle of the light cone on the cutoff surface and asymptotically in the coordinates \eqref{BTZ}. The computation is logically different from the one used to derive \eqref{propvv}, but they give the same result. }  This is  further holographic evidence that the effective signal propagation speed in the finite temperature QFT on the cutoff surface is superluminal.

\renewcommand{\e}[1]{e^{#1}}

\section{Exact Holographic RG}\label{sec:RG}

In this section we present some more details about the relationship between the $T \bar T$ deformed theory and the holographic RG. 
On the QFT side, we use the Zamolodchikov equation to derive an exact RG equation for the scale dependence of the partition function.
We then show that this RG equation is identical to the Hamilton-Jacobi equation that expresses the dependence of the bulk gravity action with a radial cutoff on the radial location of the boundary.  Finally, we present a more precise definition of the $T \bar T$ deformed theory in terms of a suitable Hubbard-Stratonovich transformation, which has been shown to act as an intertwining map between the Weyl anomaly equation of a 2D CFT partition function and the Wheeler-DeWitt equation in 3D gravity \cite{HV,freidel}. This correspondence further substantiates our interpretation of the coupling constant $\mu$  as the radial location in the bulk.

\subsection{Zamolodchikov and Wilson-Polchinski}

The $T \bar T$ deformed conformal field theories are interesting and special, because they allow for an exact study of their renormalization group flow. The $T\bar T$ interaction term introduces an effective UV cutoff scale, parametrized by the irrelevant coupling $\mu$. Hence if we consider the deformed CFT on a 2D  space-time with arbitrary metric $ds^2 = g_{\alpha\beta}\spc dx^\alpha dx^\beta$, the partition function and all other quantities will acquire a non-trivial dependence under Weyl rescalings. The goal of the exact renormalization group is to give a complete description of this scale dependence. 

Metric variations of QFT quantities are governed by the action of the stress-energy tensor. Variations of the Weyl factor are generated, on the one hand, by inserting the trace of the stress tensor $\Theta =\frac14 T^\alpha_\alpha$, and on the other hand by varying the $T \bar T$ coupling $\mu$, which amounts to an insertion of the composite operator $T \bar T$. Our strategy is to write this relation, in combination with the Zamolodchikov equation, in the form of an exact RG equation. Here we will work to leading order in a derivative expansion,
a more complete presentation is given in subsection \ref{hstransform}.

To extract the behavior of the partition function under Weyl transformations, we separate out the scale factor, and parametrize the 2D metric via
  \es{weyldeco}{
ds^2 & = e^{2\varphi(x)}\,  \hat{g}_{\alpha\beta} dx^\alpha dx^\beta\,,
}
where $\hat{g}_{\alpha\beta}$ specifies a unit determinant metric. By definition we have
 \begin{align}\label{thetadef}
 	\corr{\spc \Theta\spc } &\equiv -{e^{-2\varphi}\ov 4}\,\deltadelta{\log Z_\text{QFT}}{\varphi}\,.
 \end{align}
The undeformed CFT partition function transforms as $Z_\text{CFT}(g) = e^{A(\varphi,\hat{g})}Z_\text{CFT}(\hat{g})$
where the prefactor $e^{A(\varphi,\hat{g})}$ accounts for the scale dependence due to the trace anomaly 
\es{traceanomaly}{
{\delta A(\varphi,\hat{g}) \ov \delta \varphi} =  -{c \ov 24 \pi}e^{-2\varphi}\spc R(\varphi, \hat{g})\,,
}
where $c$ is the central charge of the CFT. Putting this together with \eqref{thetadef}, the trace anomaly $\bigl\< \Theta \bigr \> =-{c\ov 96\pi}\, R(\varphi,\hat{g})$ in a CFT follows. 

One can easily establish to first order in conformal perturbation theory that in flat space in the deformed theory:
\es{ThetaExact}{
\bigl\< \Theta \bigr \> &=- {\mu\ov 2}  \bigl \< T \bar T\bigr\>\,.
}
It was found in~\cite{italians}, for the special case of free scalars treated as a classical field theory, that this equation holds to all orders in $\mu$. We will assume that this is true for any deformed CFT. 

Motivated by combining the trace anomaly \eqref{traceanomaly} and \eqref{ThetaExact}, we conjecture that for holographic CFTs
\es{FuncDer}{
\bigl\< \Theta \bigr \> &=-{c\ov 96\pi}\, R(\varphi,\hat{g})- {\mu\ov 2}  \bigl \< T \bar T\bigr\>\,,
}
where $T \bar T$ is short-hand for the combination
\es{ttbshort}{ 
T \bar T & = \frac 1 8 \sqrt{g}\, g^{\al\ga}g^{\beta\de}\, \Bigl(T_{\al\beta}T_{\ga\de}  -   T_{\al\ga} T_{\beta\de}\Bigr)\,,
}
together with the Zamolodchikov relation \eqref{zamttb} holds for an arbitrary $\varphi$. Then we can factorize $ \< T \bar T\bigr\>$ to obtain for $\hat{g}_{\alpha\beta} = \eta_{\alpha\beta}$,
\es{CFT-HJ}{
 	\corr{\Theta} &= -\frac{c}{96\pi} \spc R(\varphi) - {\mu\ov2}  \bigr(\corr{T}\corr{\bar{T}}  -   \corr{\Theta}\corr{\Theta} \bigl)\,.
}
This is our proposed form of the exact renormalization group equation of the $T \bar T$ deformed CFT. It can be shown to hold exactly for hyperbolic (and flat) space, hence for slowly varying  $\varphi$ it should be regarded as the leading term in a derivative expansion. Below, we will provide concrete evidence that for holographic CFTs, the subleading terms in the derivative expansion are negligible, and \eqref{CFT-HJ} continues to hold at finite values of $\mu$ and arbitrary $\hat{g}_{\alpha\beta}$. This is the main result of this section.

Suppose we define an effective action $S_\text{cl}(g,\mu)$ via\footnote{ Here the normalization is chosen with an eye towards gravity where the prefactor would equal $1/16\pi \GN$. }
\es{EffAction}{
Z_\text{QFT}(g, \mu)&\equiv \exp\le(- {c\ov 24\pi} S_\text{cl}(g,\mu)\ri)\, .
}
With this definition, we can rewrite equation \eqref{CFT-HJ} as\footnote{
We start the rewriting of \eqref{CFT-HJ} by decomposing the stress tensor and metric variations into traceless and trace parts $T_{\al\beta} =\hat{T}_{\al\beta}+{g_{\al\beta}\ov2}\, T^\ga_\ga,$ and $\de g^{\al\beta} =e^{-2\varphi}\le(\de \hat{g}^{\al\beta}-2\hat{g}^{\al\beta}\,\de\varphi  \ri)\,.$ This leads to the relation:
$$
\de\log Z_\text{QFT}\spc =\spc \int d^2x \ \sqrt{g} \le[\frac12\, e^{-2\varphi}\, \hat{T}_{\al\beta} \spc  \de \hat{g}^{\al\beta}-T^\al_\al \de\varphi\ri]\,.
$$
Equation \eqref{hjeqo}  follows by combining this relation with the fact that 
$$ 
\corr{T}\corr{\bar{T}} \spc =\spc \textstyle  \frac 1 8   g^{\al\ga}g^{\beta\de}\, \bigl\<\hat{T}_{\al\beta}\bigr\>\big\<\hat{T}_{\ga\de}\bigr\>\,.
$$
}
\es{hjeqo}{
{\delta  S_\text{cl}\ov \delta \varphi }=-e^{2\varphi}\spc R(\varphi,\hat{g})- {c\spc \mu \ov 24\pi}\spc e^{-2\varphi}  
\le(\hat{g}^{\al\ga}\hat{g}^{\beta\de}\,{\delta  S_\text{cl}\ov \delta \hat{g}^{\alpha\beta} }\,{\delta  S_\text{cl}\ov \delta \hat{g}^{\ga\de} }-{1\ov 8} \le({\delta  S_\text{cl}\ov \delta \varphi } \ri)^2\ri)\,.
}
This an exact flow equation for the effective action $S_\text{cl}(g,\mu)$ analogous to the Wilson-Polchinski exact RG equation. It has been recognized for some time \cite{boer-vv}
as we will now show, this equation precisely agrees the Hamilton-Jacobi equation that describes the radial dependence of the classical action in 3D gravity on an AdS space-time with a radial cutoff, provided we set $\mu = {24\pi \ov c}$.

\subsection{WDW and Hamilton-Jacobi}

We give a brief review of the holographic RG and its relation with the Wheeler-DeWitt and Hamilton-Jacobi equations. As we have done throughout this paper, we will concentrate on the dynamics of the bulk metric only, and assume that all other bulk matter fields are in their vacuum configuration and do not contribute any stress-energy or higher curvature corrections. Hence we will assume that the  bulk is describe by pure Einstein gravity. 

The holographic correspondence relates renormalization group flow in the CFT to radial evolution in the AdS space-time.
The idea is to describe this evolution via a Hamiltonian formalism in which the radial direction plays the role of a euclidean time.
We start by writing the 3D metric in the ADM parametrization corresponding to a foliation of the 3D manifold by constant $r$ slices
\es{adm}{
 	ds^2 = N^2 dr_c^2 + g_{\alpha\beta}\spc \bigl(dx^\alpha + N^\alpha dr\bigr) \bigl(dx^\beta + N^\beta dr\bigr)\,
 }
Here $N$ denotes the lapse, $g_{\alpha\beta}$ the metric on a radial slice, and $N^\alpha$ the shift vector.  
Next we write  the 3D Einstein action, including the boundary action given in \eqref{gaction}, in the ADM decomposition
\es{sgrav}{
	S_\text{grav} = \int d^3 x \lrpar{\pi^{\alpha\beta}\dot{g}_{\alpha\beta} - N^\alpha \hamgrav_\alpha - N \hamgrav}.
 }
 where the dot indicates derivative with respect to the radial coordinate $r$, $H_\beta = 2 \nabla^\alpha \pi_{\alpha\beta}$ are the generators of 2D diffeomorphisms along the slice, 
\es{hwdwz}{
\hamgrav
 &  = 
2 \pi^\alpha_\alpha + 
{1 \ov \sqrt{g}} \le(\pi^{\alpha\beta}\pi_{\alpha\beta}- (\pi^\alpha_\alpha)^2 
\ri) - \sqrt{g}\spc R }
denotes the ADM Hamiltonian, and $R$ is the two dimensional Ricci scalar of $g_{\al\beta}$.\footnote
{
This form of the ADM Hamiltonian follows from the more standard expression
\es{hwdwo}{
H_\text{ADM} & = \frac{1}{\sqrt{g}} \bigl(\tilde\pi^{\alpha\gamma}\tilde\pi_{\alpha\beta} - (\tilde\pi^{\alpha}_\alpha)^2\bigr) - \sqrt{g}\lrpar{R+2}
} 
via the replacement $ \tilde\pi_{\alpha\beta} \spc =\spc  \pi_{\alpha\beta} -\sqrt{g} \spc g_{\alpha\beta}$. This shift  incorporates the extra boundary cosmological constant in \eqref{gaction}, and is designed to cancel out the constant vacuum energy term while replacing it by a linear term proportional to $\pi^\alpha_\alpha$. This is a key step in the holographic renormalization procedure and for rewriting the radial evolution as an RG flow.} 
Here the variable $\pi_{\alpha\beta}$ denotes the canonically conjugate variable to the metric $g_{\alpha\beta}$, and
$R$ is the scalar curvature on the radial slice. 
The shift and lapse  functions  are Lagrange multipliers enforcing the  the momentum and Hamiltonian   constraints   $\hamgrav_\alpha = H=0$.

Now let us define $S_\text{cl}(g)$ as the value of the total 3D action $S_\text{grav}$ evaluated on the classical background geometry with boundary values at $r=\rrc$ given by 
\es{bcmetric}{
ds^2\vert_{r=r_c} & = g_{\alpha\beta}(x) dx^\alpha dx^\beta\,. 
}
The boundary values of all other bulk fields besides the metric are set to zero. We assume that their bulk dynamics can be consistently  decoupled from the bulk dynamics of the metric.\footnote{Note that, without loss of generality, we have set $r_c =1$ compared to equation \eqref{adsmet}. The radial AdS direction is uniquely parametrized by the Weyl factor of the metric.}

 The Hamilton-Jacobi equation is a functional differential equation that governs how the on-shell value of the bulk action $S_\text{cl}(g)$, defined as in \eqref{gaction},  depends on the boundary value of the metric. It can most easily be derived by first consideingr the semi-classical partition function of the bulk theory with the same given boundary conditions. In the saddle point approximation
\es{Zaction}{
Z_\text{grav}(g)& = \exp\le(-{{1\ov 16\pi G} \, S_\text{cl}(g)}\ri)\, .
}
By letting the radial direction play the role of time, we are led to interpret this partition function as a wave-functional of the boundary metric $g$.
Accordingly, it must solve the gravitational analogue of the Schr\"odinger equation, commonly known as the Wheeler-DeWitt constraint
\es{wdweq}{
H_\text{wdw} Z_\text{grav}(g) & = 0\,,
}
where $H_\text{wdw}$ defines a functional differential operator, given by replacing in the classical ADM Hamilton \eqref{hwdwz} the momentum variables $\pi^{\alpha\beta}$ by the functional derivative with respect to the metric
\es{pideriv}{
\pi^{\alpha\beta} =  -{1\ov  \kappa } {\delta\ \ov \delta g_{\alpha\beta}}, \qquad \qquad \kappa \equiv \frac{1}{16\pi \GN}\, .
}
The WDW constraint \eqref{wdweq}  reduces to the Hamilton-Jacobi equation upon inserting \eqref{Zaction} and taking the limit $\kappa \to \infty$.

Let us  again separate out the scale factor and parametrize the 2D metric as in  \eqref{weyldeco}.
 Since $\varphi$ is a monotonic function of the radial coordinate $r$, it serves as a good parametrization for the bulk radial direction. In terms of these variables, the WDW Hamiltonian \eqref{hwdwz} takes the form\footnote{Here we use the decomposition $\pi_{\al\beta} =e^{2\varphi}  \le(\hat{\pi}_{\al\beta}+\frac12\hat{g}_{\al\beta}\spc \pi\ri)$  with
 $\pi = \pi^{\al}_\al$\,. }
\es{nhwdw}{
\spc{H}_\text{wdw} & =2 \pi + {\e{-2\varphi}}\le( \hat{\pi}^{\alpha\beta} {\hat \pi}_{\alpha\beta} - \frac 1 2 \pi^2 \ri) - {\e{2\varphi}}\spc R\,,\\[2mm]
	\hat{\pi}_{\alpha\beta}  & = -{1\ov  \kappa } {\delta\ \ov \delta \hat{g}^{\alpha\beta}} \qquad \qquad \pi = {1\ov 2  \kappa } {\delta\ \ov \delta \varphi}\,,
}
where indices are now contracted with $\hat{g}$.
Note the factor of $-1/2$ in the equation relating $\pi$ to $\de/\de \varphi$, which can be easily determined by acting on $\sqrt{g}$ with both operators.
Inserting \eqref{Zaction} into \eqref{wdweq} and taking the $\kappa\to 0$ limit yields the HJ-equation
\es{HJeqn}{
{\delta  S_\text{cl}\ov \delta \varphi } +  {e^{-2\varphi}} \le(\hat{g}^{\al\ga}\hat{g}^{\beta\de}\,{\delta  S_\text{cl}\ov \delta \hat{g}^{\alpha\beta} }\,{\delta  S_\text{cl}\ov \delta \hat{g}^{\ga\de} }-{1\ov 8} \le({\delta  S_\text{cl}\ov \delta \varphi } \ri)^2\ri)  \spc + {\e{2\varphi}}\spc R(\hat{g},\varphi) & = \spc 0\, .
}
Via holography, this equation acquires the meaning of an exact RG equation for the effective action of a CFT with a UV cutoff, defined by integrating out all CFT degrees of freedom above a scale associated to 
the value of the Weyl factor $e^\varphi$.  Up to now, however, it has not been clear what this holographic UV cutoff exactly looks like from the QFT perspective.
The precise match with the exact RG equation \eqref{CFT-HJ}-\eqref{hjeqo} is strong evidence that this preferred holographic UV cutoff is given by the $T \bar T$ deformation.

\subsection{WDW from Hubbard-Stratonovich}\label{hstransform}

The above results all have the following common geometric origin. Early studies of the modular geometry of the conformal block in 2D CFT revealed a deep connection with  quantum states of 3D gravity. Based on this, it has been known for some time that the partition function of a 2D CFT can be mapped to a solution of the WDW equation of 3D gravity via an integral transform \cite{HV,freidel}. From the CFT perspective, this transform looks like a  $T \bar T$ deformation, rewritten  in terms of a Gaussian integral over metric fluctuations.

To write the integral transform, it is convenient to parametrize the metric by means of a zweibein $e^a = e^a_\alpha dx^\alpha$ via  $g_{\alpha\beta} = \delta_{ab} \spc e^a_\alpha e^b_\beta $. The CFT partition function $Z_\text{CFT}(e)$ is a reparametrization and local Lorentz invariant  functional, with scale dependence fixed by the trace anomaly. Suppose we now define the $T \bar T$ deformed theory 
such that its partition function $Z_\text{QFT}(e)$ is obtained from the CFT partition function via (c.f. equation \eqref{randmetric})
\es{legendre}{
Z_\text{QFT}(e)  & = \int \! {\cal D}\!\spc f \;  e^{{2 \ov \mu} \! \int \!  
f^+\! \wedge \spc f^-}
\, Z_\text{CFT}(e+\! \spc f) \,.
}
It is not self-evident that this definition of the deformed CFT is  equivalent to the one we used so far, ${dS^{(\mu)}_\text{QFT}/ d\mu}=\int \! d^2x\ (T\spc \bar T)_\mu$,. However, as we show below, they both lead to the same detailed match with 3D gravity, which suggests that the two definitions do coincide for holographic CFTs with large central charge.
It was shown by Freidel in \cite{freidel}, based on earlier work \cite{HV}, that this integral transform acts like an intertwining map between the trace anomaly and the WDW equation
\es{freidel}{
\Bigl({\delta\ \ov \delta \varphi} & - {c \ov 24\pi}\spc e^{-2\varphi}\spc R(g)\Bigr)\spc Z_\text{CFT}(g) = 0
\qquad  \implies  \qquad \ 
H_\text{wdw} \, Z_\text{QFT}(g)   = 0\,,
}
where $H_\text{wdw}$ given in equation \eqref{hwdwz} and \eqref{pideriv}, and where the central charge $c$ and the Newton constant are related to  $\mu$ via \cite{freidel}
\es{relations}{
c & = 1 + {24\pi \ov \mu}\,, \qquad \qquad {1\ov 16\pi \GN} \equiv \kappa = \mu^{-1}\,.
}
 The constraint equations \eqref{freidel} both hold locally at every point in 2D space-time.

The integral transform \eqref{legendre} can be transferred to inside the CFT functional integral. This yields the following formula for the action of the deformed theory 
\es{hstrafo}{
S_\text{QFT}(e)& = \raisebox{-6pt}{$\raisebox{2pt}{min} \atop {f}$}  \le(  S_\text{CFT}(e + f) 
 \, -\, {2 \ov \mu} \int \! f^+\!\!\spc  \wedge f^- \ri)\,.
}
Given that for small fluctuations $S_\text{CFT}(e + f) = S_\text{CFT}(e) - \int \bigl( f^{+}_{\alpha}\,  T^{\spc \alpha}_+ + f^{-}_{\alpha}\,  T^{\spc \alpha}_-\bigr)$ with $T^{\spc \alpha}_\pm  = T^{\alpha}{\!}_{\beta} e_{\pm}^\beta =-{1\ov e}\spc {\delta S_\text{CFT} \ov \delta e^\pm_\alpha}$, this looks like a Hubbard-Stratonovich representation described in section \ref{speed-hs} of the $T \bar T$ deformed theory \eqref{ttbardeform} as a Gaussian integral over a fluctuating metric with deformation parameter $\mu$.
In the large $c$ limit, the integral transform \eqref{legendre} can be performed via a semi-classical approximation, and thus amounts to performing  a Legendre transformation. For small stress-energy fluctuations, the formula \eqref{hstrafo} for the QFT action then reduces to $S_\text{QFT} = S_\text{CFT} + {\mu} \int\! d^2 x\, T\spc\bar T$.

In the semi-classical large $c$ limit, Freidel's result amounts to the statement that
\es{einstein-hs}{
 S_\text{grav}(e) & =   \raisebox{-6pt}{$\raisebox{2pt}{min} \atop {f}$}  \le( S_\text{CFT}(e + f) 
 \, -\, {2\kappa} \int \! f^+\!\!\spc  \wedge f^-\ri) \,.
}
where $S_\text{grav}(e) = {\kappa} S_\text{cl}(e)$ is the classical bulk gravity action with boundary conditions $g_{\alpha\beta} = \delta_{ab} \spc e^a_\alpha e^b_\beta $ at the cutoff surface $r = r_c$. The derivation of this universal result relies on the fact that the metric dependence of the CFT action is fixed by the conformal anomaly and given by the Polyakov action
\es{polyakov}{
S_\text{CFT}(e) =  {c \ov 192\pi} \int \! d^2x \, R \, \square^{-1} R \, .
}
The precise equality
\es{finalidentity}{
S_\text{grav}(g) = S_\text{QFT}(g)
}
 between the bulk gravity action \eqref{einstein-hs} and the  metric dependence of the  effective action \eqref{hstrafo} of the deformed CFT  guarantees that all correlation functions of the stress-energy tensor in the deformed 2D CFT exactly match with those obtained from holography. The result \eqref{finalidentity} generalizes the known match between the holographic and CFT conformal anomalies \cite{skenderis} to the new situation, where the boundary is  placed at a finite distance from the center of the bulk.

We refer to  \cite{freidel} for a detailed derivation of the result \eqref{freidel}. Here we just add a short comment about its underlying intuition. 
 The integral transform \eqref{freidel} has a geometrical significance as an gluing operation that combines together the wave-functions of two chiral gravity theories (given by the chiral conformal blocks of the CFT) into a single non-chiral wave-function (given by the partition function $Z_\text{QFT}$). It has long been known \cite{HV} that the conformal Ward identities satisfied by a chiral conformal block in 2D CFT are identical to the physical state conditions on wave-functions of chiral gravity, provided one uses a holomorphic polarization on its phase space  -- that is, provided one writes the chiral wave-function in terms of complex variables  analogous to the coherent state basis of a harmonic oscillator.  In short, CFT conformal blocks, when viewed as functionals of the corresponding chiral zweibein $e^+$ (or $e^-$) are coherent states of chiral 3D gravity \cite{HV}. 
Gluing the chiral coherent state wave-functions together into a real solution of the non-chiral WDW equation requires performing an integral transform, analogous to the integral transform that rewrites the coherent state basis of a harmonic oscillator into a wave-function in the position representation. In the first order formulation of 3D gravity, this integral takes the form of a Gaussian integral given in \eqref{freidel}.

The integral transform \eqref{freidel} gives a well-controlled definition of the $T \bar T$ deformed CFT at large central charge.
 With this definition, the result by Freidel provides an independent derivation of the Zamolodchikov formula \eqref{zamttb} and the corresponding exact RG equation.
 Moreover, it implies that for large $c$, the all $n$-point connected correlation functions of the stress tensor computed in the QFT are identical to the correlation functions computed via pure 3D gravity
\begin{eqnarray}
{ (-2)^n\ov \sqrt{g_1}\spc \sqrt{g_2} \ldots \sqrt{g_n}} \, {{\de S_\text{grav}}(g) \ov {\de g^{\alpha\beta}_1 \de g^{\beta \gamma}_2 \ldots \de g^{\rho\sigma}_2}}& = & \Bigl\< T_{\alpha\beta}(x_1) T_{\beta \gamma}(x_2) \dots T_{\rho\sigma}(x_n)\Bigr\>^\text{conn}_\text{QFT(g)} \nonumber
\end{eqnarray}
with $g^{\alpha\beta}_i = g^{\al\beta}(x_i)$, etc.
This relation looks perhaps more miraculous than it really is. The right-hand side is fixed by the conformal anomaly and Ward identities, and depends only on one single dimensionless number,  the central charge $c$. Similarly, the correlation functions of boundary gravitons in 3D gravity are  fixed by the AdS analogue of soft-graviton theorems and only depend on the ratio of the AdS scale and the Planck scale. Still, this result is a useful extension of the standard AdS/CFT dictionary, that may open up new ways of probing the gravitational bulk physics.

  \section{Conclusion}\label{conclusions}

In this paper, we studied the class of 2D effective QFTs defined by turning on an irrelevant $T \bar T$ deformation in a general 2D CFT. We proposed that in the holographic dual,
the deformation corresponds to introducing a rigid cutoff surface that imposes Dirichlet boundary conditions at a finite radial location $r = \rrc$ in the bulk. As a check of the duality,
we have shown that the energy spectrum, thermodynamic properties, propagation speeds, and the metric dependence of the partition function agree on both sides. This correspondence is largely explained by the precise identification between the 2D conformal Ward identities and the physical state conditions in 3D gravity.

There are many open questions. It will be important to establish whether the $T \bar T$ deformation indeed produces a well defined unitary quantum system.
We have seen that for CFTs with $c=24$, the deformation is equivalent to the Nambu-Goto formulation of the string worldsheet theory on some general target space.
The NG theory is  soluble and appears to be a well defined deformation of the CFT for both  choices of sign, including the one that leads to superluminal propagation speeds relative to the non-dynamical background metric.  This indicates that the $T \bar T$ deformation is also consistent for large $c$ CFTs, but a general proof is not yet available.

It is natural to ask whether some of our results can be extended to higher dimension. The main catalyst our story, the Zamolodchikov equation \eqref{zamttb},  looks like a large $N$ factorization property.  So it seems plausible that an analogous equation can be derived in large $N$ CFTs in higher dimensions. However, since conformal symmetry is less restrictive for $d>2$, it is not clear if such an equation can be used to derive analogous unique flow equations for the energy levels and the partition function. Even so, it would be instructive to explore what double trace technology can teach us about the $T\bar{T}$ deformed theory. 
Because the stress tensor is normalized such that its two-point function is $\<T\, T\>_\text{CFT}=O(N^2)$, in order for the double trace coupling $\mu$ to appreciably influence the dynamics, and to preserve the structure of the large-$N$ expansion, it has to be $\mu = O(1/N^2)$ or parametrically larger. 
In the regime where $r_c/\ell_\text{AdS}=O(1)$, $\mu$ is indeed of this order. To explore sub-$\ell_\text{AdS}$ scale physics we cannot rely on a  perturbative expansion in $1/N$, and non-perturbative methods are needed. We have seen that, in two dimensions and for correlation functions of the stress tensor, such non-perturbative methods are indeed available.

It was suggested in \cite{Giombi:2013yva} based on the Hubbard-Stratonovich presentation of the $T\bar{T}$ deformed theory \eqref{Cardy} that the $\mu\to\infty$ limit of large $N$ CFTs is the CFT coupled to (emergent) quantum gravity. Our interpretation of the evolution of the spectrum with $\mu$ given in \eqref{BurgersSol} implies that this theory has only very few states, which would be interesting to understand from the emergent gravity perspective.    

For most of our computations, we have restricted our attention to long distance properties of the $T \bar T$ QFT.  Indeed, it is not clear whether it is possible to define true local operators, that probe or excite the QFT at arbitrarily short distance scales. We have seen that turning on the $T \bar T$ interaction leads to fluctuations in the effective metric that grow  large in the UV. The randomness of the dynamical UV metric complicates the task of finding a precise holographic map analogous to the standard GKPW dictionary QFT and gravity observables. Still, it would be worthwhile to study the properties of localized probes in the QFT, other than stress tensors, and investigate whether is it possible
to compute correlation function at sub-AdS distances, as measured at the cutoff surface. If we define the dimensionless coupling as the ratio
\es{Last}{\nonumber
\bar\mu \spc   \equiv  \spc \mbox{dimensionless coupling}   = {\mu \ov \Delta \theta^2} = &  {24\pi \ov c \spc r_c^2}{1\ov \Delta\theta^2 } = {24\pi \ov c }{1\ov d^2}  \\[2mm]
  d  \spc  \equiv \spc \mbox{distance scale in AdS units}  \spc& = \spc r_c\spc \Delta \theta \,,
}
we see that the $T \bar T$ interaction and the associated metric fluctuations remain small all the way down to the short distance scale $d_\text{planck} = \sqrt{24 \pi \ov c} \ell_{\text{AdS}}$. So in this sense, we should be able to use the $T \bar T$ QFT to probe bulk physics at sub-AdS distance scales. The key questions, however, are how to extend our calculations to general operators ${\cal O}$ and how the bulk physics in this regime is affected by the presence of the cutoff surface. Assuming that the cutoff surface continues to behave like a Dirichlet wall, correlation functions at this short distance scale should behave similar correlation functions in a gravitational theory in flat space.

\section*{Acknowledgements}
We thank Vijay Balasubramanian, Clay Cordova, Xi Dong, Raphael Flauger, Tom Hartman, Bruno Le Floch, Igor Klebanov, Sung-Sik Lee, Hong Liu, and Juan Maldacena for helpful discussions and comments.
 The research of M.M. was supported in part by the U.S. Department of Energy under grant No. DE-SC0016244. The research of H.V. is supported by NSF grant PHY-1620059.

 \appendix
 
 \section{Propagation speed in general backgrounds}
 
 In this appendix we generalize the analysis   of signal propagation speed presented in section \ref{sec:gravityspeed} to a more general class of states. These resemble Ba\~nados geometries, except that the induced metric is flat on the Dirichlet wall at $\rho=\rho_c$. To impose this, we start with the metric Ansatz:
 \es{MetricAnsatz}{
 ds^2=&{d\rho^2\ov \rho^2}-\rho_c^2 \spc dx^+ dx^- \\
 &+ (\rho^2-\rho_c^2) \spc h^{(1)}(x^+,x^-)_{\al\beta}\spc dx^\al dx^\beta+\le({\rho_c^4\ov \rho^2}-{ \rho_c^2}\ri)\spc h^{(2)}(x^+,x^-)_{\al\beta}\spc dx^\al dx^\beta\,,
 }
 where we used the property of 3D gravity that expansions in $\rho$ terminate after a couple of orders. Note that setting $\rho=\rho_c$ eliminates the second line in \eqref{MetricAnsatz} and the metric on the Dirichlet wall is $ds^2\vert_{\rho_c}=-\rho_c^2 \spc dx^+ dx^- $. Plugging into Einstein's equations, we get that $h^{(1,2)}_{\al\beta}$ can be parametrized by two functions:
  \es{MetricSol}{
 h^{(1)}_{\al\beta}&= {c-1\ov 4}\spc  M_{\al\beta}\,, \quad 
 h^{(2)}_{\al\beta}= {c+1\ov 4}\spc  M_{\al\beta}\,,
 \quad
 M_{\al\beta}\equiv \begin{pmatrix}
a & - c\\
-c & b
 \end{pmatrix}\,, \quad c\equiv\sqrt{ab+1}\,,
 }
 where the functions $a(x^+,x^-),\, b(x^+,x^-)$ satisfy 
 \es{pde}{
 \p_- a+\p_+ c=0\,, \qquad\quad  \p_+ b+\p_- c=0\,, \qquad\quad c=\sqrt{ab+1}\,,
 }
 which is a set of coupled nonlinear PDEs.\footnote{It is instructive to write down the planar BTZ black hole in this parametrization, which takes the form:
 \es{BTZNewCoord}{
 ds^2=&{d\rho^2\ov \rho^2}-\rho_c^2 \spc dx^+ dx^- - {2\rho_c^6\ov (\rho_c^4-1)^2} \le[(\rho^2-\rho_c^2)+\le({1\ov \rho^2}-{1\ov \rho_c^2}\ri)\ri]  \begin{pmatrix}
dx^+ & dx^- 
 \end{pmatrix}
 \begin{pmatrix}
d & {\rho_c^4+1\ov 2\rho_c^2}\\
{\rho_c^4+1\ov 2\rho_c^2} & {1\ov d}
 \end{pmatrix}
 \begin{pmatrix}
dx^+ \\ dx^- 
 \end{pmatrix}
  }
 where we chose $\rho_c>1$, and $d>0$ is a parameter characterizing the solution. In the parametrization \eqref{MetricSol} the solution corresponds to
 \es{abValue}{
 a&= - {2\rho_c^2\ov \rho_c^4-1} \spc d\,, \qquad b= -{2\rho_c^2\ov \rho_c^4-1} \spc {1\ov d}\,.
 }}
 These can be solved in a series form:
 \es{SerSol}{
a(x^+,x^-)&=\ep  \spc A'(x^+)- {\ep^2\ov 2}A''(x^+)B(x^-)+O(\ep^3)\\
b(x^+,x^-)&=\ep  \spc B'(x^-)- {\ep^2\ov 2}A(x^+)B''(x^-)+O(\ep^3)\,,
 }
 hence $c=O(\ep^2)$.  Of course, this is locally just AdS$_3$ in complicated coordinates.
 
 Now consider a fluctuation in the location of the Dirichlet wall  $\rho_c \to \rho_c + \delta \rho(x^+,x^-)$, and require that the resulting metric stays flat
 \es{Ricci}{
0&=R\le(\rho_c + \delta \rho(t,\theta)\ri)= -{8\ov r_c^3}\le[\p_+\spc\p_-+{\ep\ov 2}\le(A'(x^+)\spc \p_-^2+B'(x^-)\spc \p_+^2\ri)\ri]\delta \rho(x^+,x^-) +O(\ep^2,\delta \rho^2)\,.
 }
 Assuming that $A'(x^+)$ and $B'(x^-)$ are slowly varying, we get the corrected propagation speeds to be:
 \es{vpmSpace}{
 v_+=1-\ep\spc B'(x^-)\,,\qquad \qquad  v_-=1-\ep\spc A'(x^+)\,.
 }
 To $O(\ep)$ the metric \eqref{MetricAnsatz} takes the form
  \es{MetricSol2}{
 ds^2=&{d\rho^2\ov \rho^2}-\rho_c^2 \spc dx^+ dx^- +\le({\rho_c^4\ov \rho^2}-{ \rho_c^2}\ri)\spc \le({\ep\ov2}  \spc A'(x^+)(dx^+)^2+{\ep\ov2}  \spc B'(x^-)(dx^-)^2\ri)+O(\ep^2)\,.
 }
  From the behavior of the metric near $r=r_c$ and the usual definition of the holographic stress tensor \cite{holostress}, we deduce the following expectation values in the dual field theory:
\es{Texp}{
\bigl\<T_{++}(x^+)\bigr\> =-{\rho_c^2\ov 16\pi \GN}\, \ep \spc A'(x^+), \qquad\  \bigl\< T_{--}(x^-)\bigr\>=-{\rho_c^2\ov 16\pi \GN}\, \ep \spc B'(x^-)\,.
} 
  We combine  this equation with \eqref{vpmSpace} to obtain \eqref{LCAdS2}. We note that there is an intriguing connection between this equation and the  Nambu-Goto string: if we rename $X^-\equiv-\ep B,\, X^-\equiv-\ep A$, we obtain the Virasoro conditions \eqref{virconst}.
  
    \bibliographystyle{ssg}
  \bibliography{TTbar}

 \end{document}